\documentclass[twocolumn]{article}
\usepackage[margin=16mm]{geometry}
\usepackage{sectsty}
\sectionfont{\normalsize}
\subsectionfont{\normalsize}
\usepackage{authblk}
\usepackage{cite}
\usepackage{lineno}
\usepackage{siunitx}
\usepackage{amsmath}
\usepackage{amsfonts}
\usepackage{amssymb}
\usepackage{graphicx}
\usepackage[font=small]{caption}
\usepackage{hyperref}
\hypersetup{
    colorlinks=true,
    linkcolor=blue,
    filecolor=magenta,      
    urlcolor=cyan,
    }

\providecommand{\keywords}[1]
{
  \small	
  \textbf{\textit{Keywords---}} #1
}

\title{
High-speed photoelastic tomography for axisymmetric stress fields in a soft material: temporal evolution of all stress components
}

\author[1]{Yuto Yokoyama}
\author[1]{Sayaka Ichihara}
\author[1]{Yoshiyuki Tagawa}

\affil[1]{Department of Mechanical Systems Engineering, Tokyo University of Agriculture and Technology, Koganei, Tokyo 1848588, Japan
}

\date{}

\begin{document}

\twocolumn[

\maketitle

\begin{abstract}
This study presents a novel approach for reconstructing all stress components of the dynamic axisymmetric fields of a soft material using photoelastic tomography (PT) and a high-speed polarization camera. 
This study focuses on static and dynamic Hertzian contact as an example of transient stress field reconstructions.
For the static Hertzian contact (a solid sphere pressed against a gel block), all stress components in the urethane gel, which has an elastic modulus of 47.4 kPa, were reconstructed by PT using the measured photoelastic parameters. The results were compared with theoretical solutions and showed good agreement.
For the dynamic Hertzian contact (a sphere impacting gel), a high-speed polarization camera was used to reconstruct the transient stress field within the gel.
PT was used to quantitatively measure the shear and axial stress waves and showed different propagation speeds on the substrate.
The technique allowed the simultaneous measurement of stress fields ranging from $\mathcal{O}(10^{-1})$ to $\mathcal{O}(10^1)$ kPa during large deformations, demonstrating its accuracy in capturing rapidly changing stress tensor components in dynamic scenarios.
The scaling laws of the calculated impact force agreed with theoretical predictions, validating the accuracy of PT for measuring dynamic axisymmetric stress fields in soft materials.
\end{abstract}

\vspace{3mm}


\vspace{8mm}
]

\section{Introduction}

Evaluating stress in soft materials is of great importance in the fields of biomedical engineering and cell printing applications.
For example, in biomedical engineering, it is essential to determine the stress distributions around cerebral aneurysms to decipher the mechanisms leading to their rupture \cite{shojima2004,meng2014,vanooij2015}.
Investigating the stress field induced by needle-free injection for evaluating pain levels also requires stress measurements in the tissue, which is also a soft material \cite{tagawa2013a,kiyama2019a}.
Moreover, understanding the stress distribution within a material resulting from droplet or liquid jet impact is pertinent to diverse engineering processes, making it a topic of considerable interest for research \cite{mitchell2019,sun2022,mitchell2019a,gordillo2018,cheng2022}.

Photoelasticity is a well-known technique for estimating stress distributions within materials \cite{frocht1941,aben1993a,asai2019,ramesh2016} and is expected to be useful in the scenarios mentioned above.
Photoelasticity operates on the principle of birefringence in materials.
Birefringence is an optical anisotropy, in which the local refractive index of a stressed material changes in different directions.
By observing the intensity of polarized light passing through the stressed body, one can measure the difference between two mutually orthogonal refractive indices and the orientation of its axis to estimate the material's internal stress state.
The stress-optic law establishes a proportional relationship between the integration of the secondary principal stress difference $\sigma_d$ in a material and the phase retardation of transmitted light $\Delta$ \cite{aben1993a,ramesh2021}, $\Delta = C \int \sigma_d(y) dy$.
Here, the coefficient $C$ --- known as the stress-optic coefficient --- is specific to the material \cite{aben1993a}, and $y$ is the camera's optical axis.
The secondary principal stress is the apparent principal stress that appears in the cross-section perpendicular to $y$-axis \cite{yokoyama2023}.
Additionally, the orientation $\phi$ of the transmitted polarized light is related to the direction of the secondary principal stress.
In this study, the retardation and orientation of the polarized light are called photoelastic parameters.

Photoelasticity possesses distinct advantages over other measurement methods, such as pressure sensors, due to its ability to measure the entire stress field within a material while being non-invasive.
Digital image correlation (DIC) \cite{hall2012,sun2022} can also be used to obtain a full-field stress measurement by analyzing local material displacements and applying the material's constitutive equation.
Such a technique requires particles to be mixed into the material as a marker to measure the displacement or optics and to produce very thin laser sheets.
In contrast, photoelasticity can obtain measurement data on stress values without contaminating the material by installing markers, as long as the stress-optic law is valid.
Additionally, the experimental setup of photoelasticity requires only circularly polarized light as a light source and a polarization camera.

Methods designed for measuring three-dimensional stress fields in materials are called integrated photoelasticity methods \cite{aben1993,aben1989,aben2000,ramesh2016}.
In this context, ``integrated'' signifies that the recorded retardation captures incremental photoelastic parameters at each material point along the path of the light ray, depending on the stress state at that specific point.
Previously, integrated photoelasticity has mainly been used for hard materials such as glass to visualize its residual stress \cite{frocht1941,aben1993a,asai2019,ramesh2016,nelson2021,yoshida2012}.
In contrast, our previous study verified the applicability of integrated photoelasticity to a soft material experiencing an axisymmetric stress field \cite{yokoyama2023,mitchell2023}.
Consequently, we have demonstrated that integrated photoelasticity is valid for a gel that is about 1,000,000 times softer than glass.

\begin{figure}[t]
    \centering
    \includegraphics[width=0.9\columnwidth]{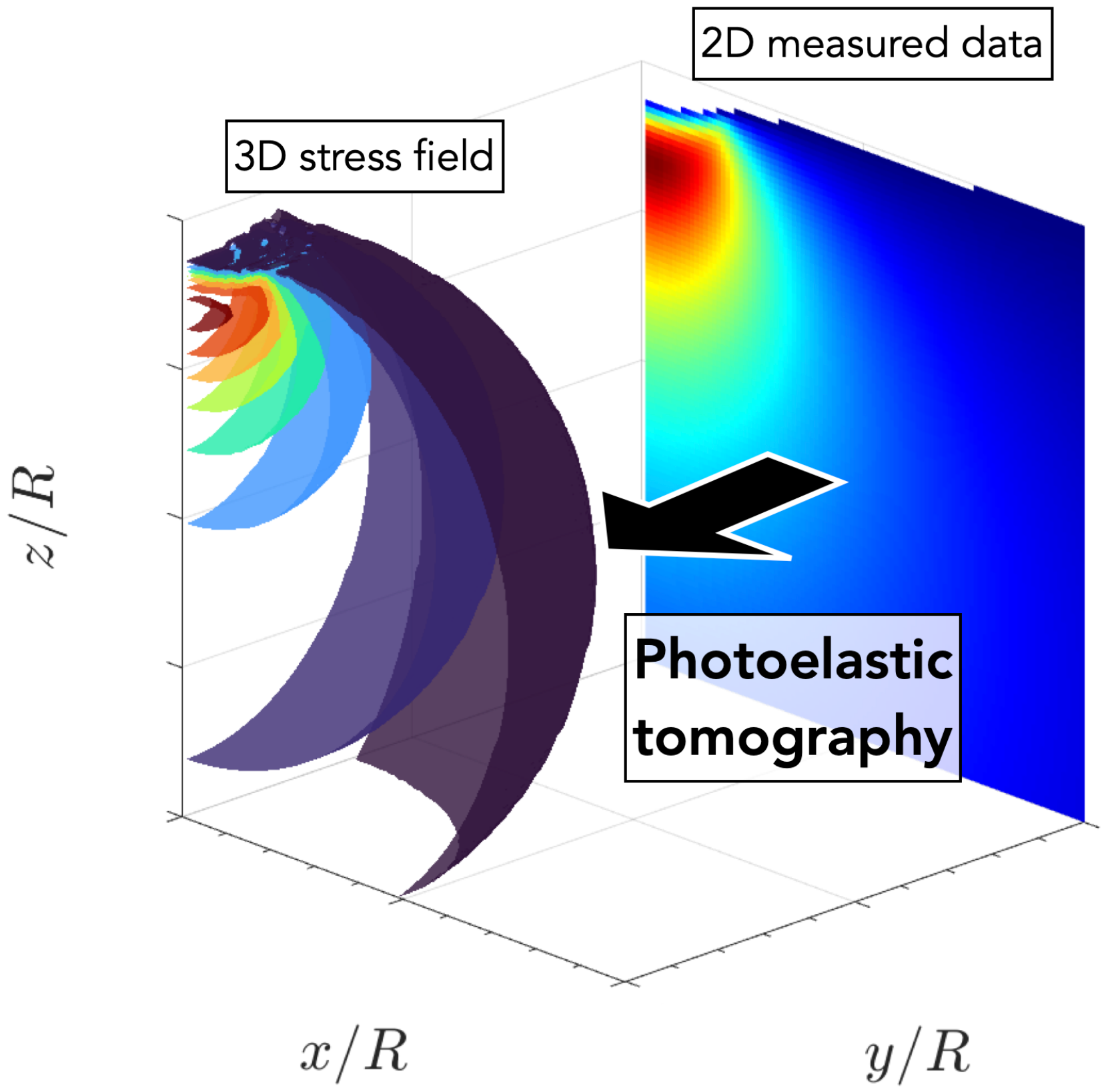}
    \caption{The relationship between the three-dimensional stress field in a material and the photoelastic parameters measured by a polarization camera.}
    \label{fig:tomography}
\end{figure}

As mentioned above, the photoelastic parameters (the phase retardation and orientation) of polarized light passing through a stressed material are related to the integrated values of the stress field inside the material, as shown in Fig. \ref{fig:tomography}. 
Therefore, it is necessary to reconstruct the internal stress field from the measured photoelastic parameters to obtain the stress field in a material.
This is called photoelastic tomography and was studied intensively by Aben {\it et al.} \cite{aben2012,aben2010a,aben2008}.
Tomography is well known as a non-destructive method for analyzing the internal structure of three-dimensional objects.
For example, computed tomography (CT) scans, which are based on X-rays, use the attenuation of the X-rays transmitted through an object to determine its internal structure.
In such cases, tomography is relatively easy because the quantities that characterize the internal structure at each point are scalar values.
On the other hand, stress is a tensor with up to six components at each point inside the material.
Two values are obtained by integrating them --- the phase retardation and orientation --- which is a small number compared to the number of tensor components to be reconstructed.
Therefore, tomography of such a field is generally difficult, but if the stress field is axisymmetric, tomography of the stress field can be performed because the number of tensor components is reduced \cite{errapart2011,anton2008}, i.e., 
\begin{equation}\label{eq:axisymmetric_stress_tensor}
    \boldsymbol{\sigma} = \begin{bmatrix}
        \sigma_{rr} & 0 & \sigma_{rz} \\
        0 & \sigma_{\theta\theta} & 0 \\
        \sigma_{rz} & 0 & \sigma_{zz}
    \end{bmatrix}.
\end{equation}
Here, $\sigma_{rr}$, $\sigma_{\theta\theta}$, $\sigma_{zz}$, and $\sigma_{rz}$ are the radial, hoop, axial, and shear stresses in cylindrical coordinates, respectively.

However, photoelastic tomography has not yet been applied to the full-field measurement of the internal stress of a soft material.
Additionally, photoelasticity has been used primarily to measure steady-state stress fields, because measuring photoelastic parameters using the phase-shifting method requires multiple images to be taken for the same stress field.
However, the development of a polarization camera with polarizers of different angles installed in the camera sensor has made it possible to measure photoelastic parameters in a single shot \cite{yoneyama2006}.
Furthermore, high-speed polarization \cite{onuma2014} can measure transient photoelastic parameters even for very high-speed phenomena.
However, although the high-speed polarization camera has been used to ``visualize'' dynamic three-dimensional stress fields (i.e., to measure photoelastic parameters) in a soft material \cite{miyazaki2021}, to the best of our knowledge, there have been no examples of quantitative measurement through the reconstruction of stresses in soft materials from the measured photoelastic parameters based on photoelastic tomography.

In the present work, photoelastic tomography is applied to measure all stress components of dynamic axisymmetric fields in a soft material using a high-speed polarization camera to address the gaps described above.
Combining photoelastic tomography and the high-speed polarization camera allows us to reconstruct all the components of the dynamic axisymmetric stress fields in a soft material.
The ability to measure all components of the stress tensor not only allows analysis of the principal stresses and their directions (rather than the secondary principal stresses) but also allows the measurement of the von Mises stress, which is used to evaluate the yield points of the material \cite{mitchell2023}.
The examination conducted in this study involves both static and dynamic Hertzian contact problems.
There are two primary reasons for this.
Firstly, if the external force and other contact conditions are known a priori, the stress field within the material or impact forces acting on the surface of the material can be obtained theoretically \cite{johnson1985,ike2019,love1929,pradipto2021,kuwabara1987}.
This allows us to evaluate the certainty of the reconstructed stress field from the experimentally obtained photoelastic parameters.
Secondly, dynamic axisymmetric stress fields appear in many phenomena, e.g., a droplet impacting a soft surface \cite{howland2016,basso2020a}, a liquid jet penetrating a gel \cite{miyazaki2021,tagawa2013a,kiyama2019a}, and cavitation generation in a gel \cite{rapet2019}.
For these two reasons, this study demonstrates that photoelastic tomography is an effective technique for measuring dynamic axisymmetric stress fields in materials.
We believe that photoelastic tomography will find broad applications as a method for understanding the high-speed phenomena of soft materials.

In Sec. \ref{sec:methodology}, the integrated photoelasticity and the photoelastic tomography algorithm are described.
The experimental method used to measure the photoelastic parameters with a high-speed polarization camera is also explained.
Section \ref{sec:ResultandDiscussion} compares the experimental and theoretical results, and Sec. \ref{sec:conclusion} summarizes the findings of this study. 

\section{Methodology}\label{sec:methodology}

\subsection{Integrated photoelasticity}

When circularly polarized light from a light source enters a material under stress, it is emitted as elliptically polarized light with photoelastic parameters (phase retardation $\Delta$ and orientation $\phi$) corresponding to the stress state in the material \cite{aben1993}.
$\Delta$ and $\phi$ can be temporally measured using a high-speed polarization camera \cite{onuma2014}.
The following relationship exists between the stress field inside a material and its photoelastic parameters \cite{aben1997,doyle1982,aben2010}:
\begin{eqnarray}\label{eq:delta-stress}
    V_1 \equiv \Delta \cos 2 \phi &=& C \int^\infty_{-\infty} \left(\sigma_{xx}-\sigma_{zz}\right) dy, \label{eq:V1}\\
    V_2 \equiv \Delta \sin 2 \phi &=& 2C \int^\infty_{-\infty} \sigma_{xz} dy,\label{eq:V2}
\end{eqnarray}
where $C$ is a material-specific value called the stress-optic coefficient and $\sigma_{xx}, \sigma_{zz}$, and $\sigma_{xz}$ are the stress components in Cartesian coordinates with the $y$-axis as the camera's optical axis.
If $\Delta$ is smaller than $\lambda/4$ and the rotation of the principal stress direction does not exceed $\pi/6$, Eqs. (\ref{eq:V1}) and (\ref{eq:V2}) are valid \cite{aben1989,aben1993a}.
Note that all of the experimental data in this study are within the bounds of this assumption.
As shown by the integrated equations, the photoelastic parameters that can be measured using a polarization camera are the integrated values of the internal stress field along the camera's optical axis.
Therefore, tomography is required to convert the photoelastic parameters into the stress tensor components at each point inside the material.
However, as mentioned in the previous section, the tomography of stress fields differs from traditional tomography, which targets a problem where a single scalar characterizes each point inside the material.
Because stress is represented by a tensor, applying conventional scalar tomography directly to a stress field is difficult.

\begin{figure}[t]
    \centering
    \includegraphics[width=1\columnwidth]{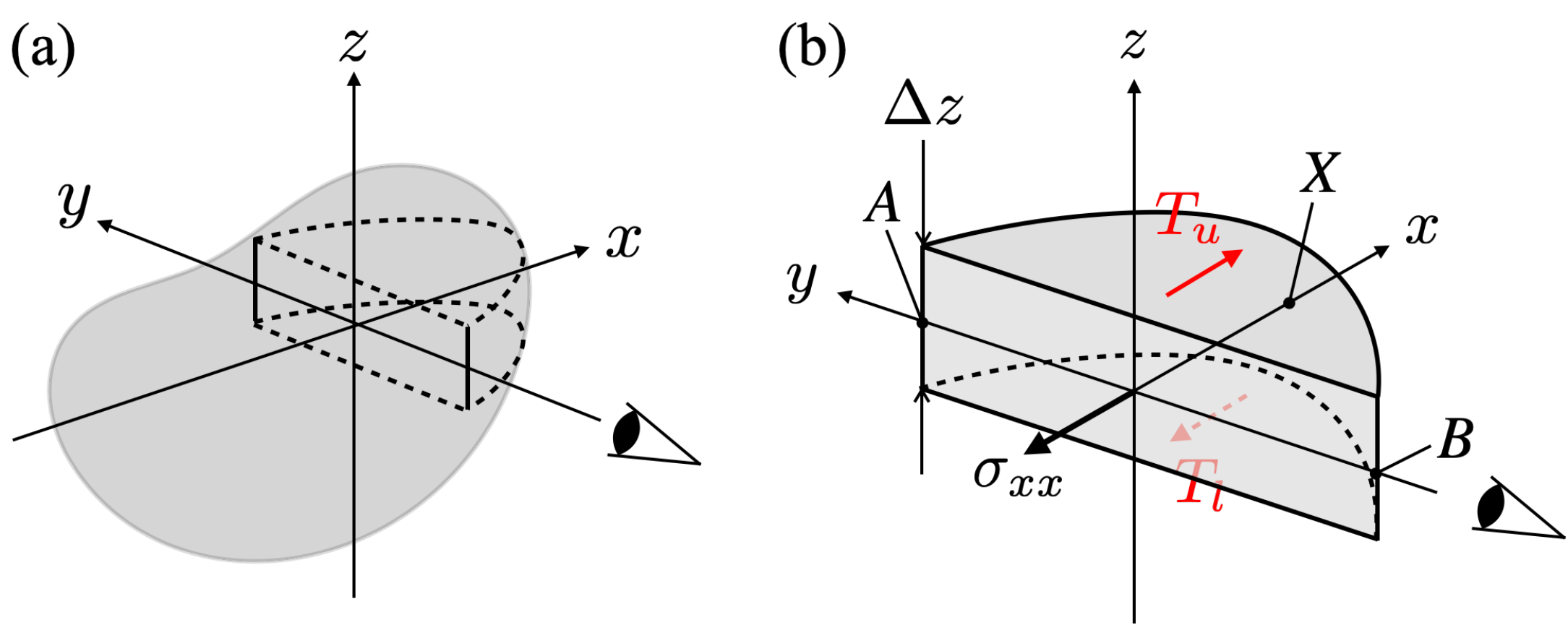}
    \caption{(a) A general three-dimensional stress field. (b) The considered element ABX, enclosed by the dashed line.}
    \label{fig:general_stress_field}
\end{figure}

Aben {\it et al.} suggested that the problem of stress field tomography could be solved if it could be reduced to a problem of scalar field tomography for a single stress tensor component \cite{aben1993,aben2012}.
To determine the stress field inside a material using photoelastic tomography, it is necessary to measure the photoelastic parameters in two sections, the upper section and lower section, parallel to the $x$-$ y$ plane and separated by a distance $\Delta z$ \cite{aben1992,anton2008}.
Generally, this measurement should be implemented for many directions of the camera's optical axis around the $z$-axis.
We now consider an arbitrary three-dimensional stress field with boundaries (Fig. \ref{fig:general_stress_field}(a)).
In a part of this field, the element ABX (see Fig. \ref{fig:general_stress_field}(b)), the force equilibrium along the $x$-axis is
\begin{align}\label{eq:equilibrium of shear force}
    \Delta z \int^A_B \sigma_{xx}dy = T_u - T_l,
\end{align}
where $T_u$ and $T_l$ are shear forces on the upper and lower surfaces of the element and $A, B$ denote the boundary of the three-dimensional stress field on the $y$-axis.
$T_u$ and $T_l$ can be described using Eq. (\ref{eq:V2}):
\begin{align}\label{eq:upper and lower shear force}
    T_u = \frac{1}{2C}\int^{X'}_{x} V'_2 dx, \quad T_l = \frac{1}{2C}\int^X_{x} V_2 dx. \\
    \because \, T_l = \int^{X}_x\int^B_A\sigma_{xz}dydx = \int^X_x\frac{1}{2C}V_2dx \nonumber
\end{align}
Here, $V_2'$ is the $V_2$ measured in the upper section, and $X'$ and $X$ are the outer positions of the upper and lower sections, respectively.
Substituting Eqs. (\ref{eq:equilibrium of shear force}) and (\ref{eq:upper and lower shear force}) into Eqs. (\ref{eq:V1}) and (\ref{eq:V2}), we obtain
\begin{align} 
    &\int^A_B\sigma_{xz}dy = \frac{1}{2C}V_2, \label{eq:tomo1}\\
    &\int^A_B\sigma_{zz}dy = \frac{1}{2C\Delta z}\left( \int^X_{x} V'_2 dx - \int^X_{x} V_2 dx \right) - \frac{V_1}{C}. \label{eq:tomo2}
\end{align}
Using this equation, the tensor tomography of the stress field can be treated as scalar field tomography.

Note that the reconstruction of the axial stress using Eq. (\ref{eq:tomo2}) requires the assumption of the equilibrium of the force, as mentioned above.
In contrast, the reconstruction of the shear stress using Eq. (\ref{eq:tomo1}) does not require this assumption.

\subsection{Photoelastic tomography}\label{sec:AlgorithmTomography}

If the stress field is axisymmetric, for example, in the problem of a rigid sphere making contact perpendicular to a flat surface, the stress field can be experimentally determined using Eqs. (\ref{eq:tomo1}) and (\ref{eq:tomo2}) \cite{anton2008,errapart2011}.
When the stress field is axisymmetric, the stress tensor has four components: radial $\sigma_{rr}$, hoop $\sigma_{\theta\theta}$, axial $\sigma_{zz}$, and shear stresses $\sigma_{rz}$, i.e., Eq. (\ref{eq:axisymmetric_stress_tensor}).
Here, the plane perpendicular to the camera's optical axis is the $r$-$z$ plane, and the axis of symmetry is the $z$-axis.
The components $\sigma_{rz}$ and $\sigma_{zz}$ are reconstructed from the measured values of $\Delta$ and $\phi$ using the integrated relationships provided in the last section.
The remaining components, $\sigma_{rr}$ and $\sigma_{\theta\theta}$, are reconstructed from $\sigma_{rz}$ and $\sigma_{zz}$ using the equations of the theory of linear elasticity \cite{errapart2011}.
In this subsection, tomography for general axisymmetric scalar fields and its algorithm will be described, followed by the application of the algorithm to tomography for stress fields.

\subsubsection{Abel transform and the onion-peeling method}

\begin{figure}[t]
    \centering
    \includegraphics[width=1\columnwidth]{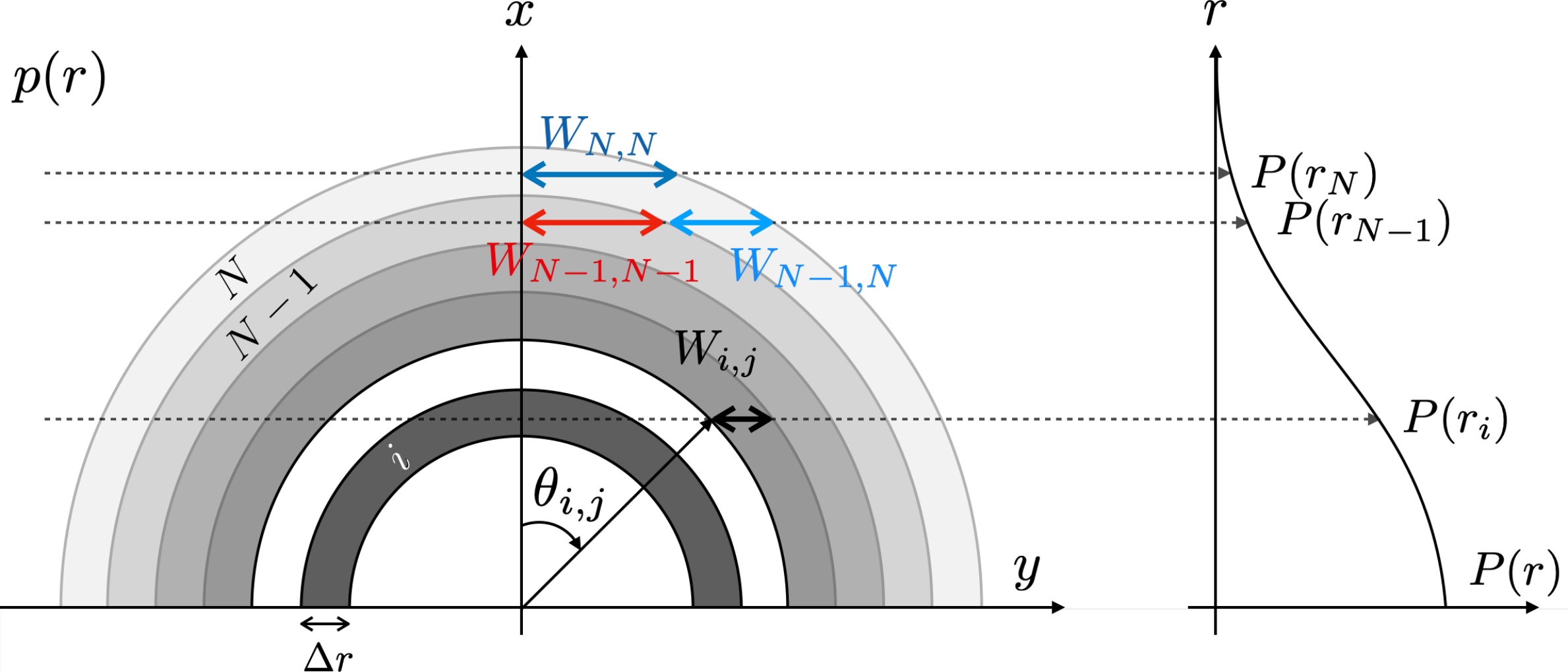}
    \caption{A schematic of the algorithm of the onion-peeling method. The relationship between the axisymmetric field $p(r)$ to be reconstructed and the projection data $P(r)$.}
    \label{fig:Onion-peeling method}
\end{figure}

A well-known reconstruction method for an axisymmetric field is the inverse Abel transform.
Generally, the relationship between an axisymmetric field $p(r)$ and a projection function $P(x)$, as shown in Fig. \ref{fig:Onion-peeling method}, which is obtained by integrating over $p(r)$, is called Abel transform and can be expressed as:
\begin{equation}
    P(x) = \int^{\infty}_{-\infty} p\left(\sqrt{x^2 + y^2}\right) dy,
\end{equation}
where $r=\sqrt{x^2 + y^2}$.
This equation can be rewritten as
\begin{equation}
    P(x) = 2\int^\infty_x \frac{p(r)r}{\sqrt{r^2 - x^2}}dr.
\end{equation}
The inverse Abel transform is given by 
\begin{equation}
    p(r) = - \frac{1}{\pi} \int^\infty_r \frac{dP}{dx}\frac{dx}{\sqrt{x^2 - r^2}}.
\end{equation}
Since the projection data obtained in the experiment are always discretized, this inverse transformation must be solved numerically.
Direct use of the above equation not only gives rise to a singularity at $x=r$, but also requires that the exact change in $P$ in the $x$ direction be known a priori.
However, the projection data obtained in the experiment are not perfectly smooth.
It should also be noted that Abel's forward and inverse transformations cover the entire space and have infinite integration limits, whereas the experimental data are limited to a finite range.
Outside of these ranges, $p$ and $P$ must be strictly zero, and any deviation will compromise the accuracy of the transform.
Basically, only local objects that are completely contained within the image frame can be accurately transformed.

The onion-peeling (OP) method is one of the numerical algorithms used to solve the inverse transform \cite{dasch1992,xiong2020a}.
The OP method considers the axisymmetric field as layers of $N$ rings (Fig. \ref{fig:Onion-peeling method}), which have a thickness of $\Delta r$ and a fixed scalar value.
The outermost ring's value $p(r_N)$ is obtained from the projection data $P(r_N)$ and the length of the optical path $2W_{N,N}$ using the equation
\begin{equation}
    P(r_N) = 2W_{N,N}p(r_N).
\end{equation}
The value of the $N$-$1$th ring $p(r_{N-1})$ can be obtained by using the projection data $P(r_{N-1})$, the value of the $N$th ring $p(r_{N-1})$ obtained earlier, and the length of the optical path for each ring, $2W_{N-1,N-1}$ and $2W_{N-1,N}$, i.e.,
\begin{equation}
    P(r_{N-1}) = 2W_{N-1,N-1}p(r_{N-1}) + 2W_{N-1,N}p(r_N).
\end{equation}
Therefore, the projection data of the $i$th ring can be described as
\begin{equation}
    P(r_i) = 2 \sum^N_{j=i} W_{i,j} p(r_j),
\end{equation}
where $r_i = i\Delta r$ is the distance from the central axis and
\begin{equation}
  W_{i,j}=
  \begin{cases}
    0 & j<i, \\
    \frac{\Delta r}{2} \sqrt{(2j+i)^2 - 4i^2}                 & j=i, \\
    \frac{\Delta r}{2}\sqrt{(2j+i)^2 - 4i^2} - \frac{\Delta x}{2}\sqrt{(2j-i)^2 - 4i^2}      & j>i.
  \end{cases}
\end{equation}

\subsubsection{Determination of the shear and axial stresses}\label{sec:Determination_o_rz_o_zz}

We determine the shear stress $\sigma_{rz}$ from the photoelastic parameters using Eq. (\ref{eq:V2}) based on the OP method.
While the OP method described earlier targets axisymmetric fields, the shear stress $\sigma_{rz}$ is no longer axisymmetric when it is distributed in the $x$-$y$ plane in Cartesian coordinates.
The relationship between the shear stresses in Cartesian coordinates $\sigma_{xz}$ and cylindrical coordinates $\sigma_{rz}$ is
\begin{equation}
    \sigma_{xz} = \sigma_{rz} \cos \theta. \label{eq:rotation}
\end{equation}
Therefore, the angle $\theta$ has to be considered when we relate the integrated value of the photoelastic parameters and the shear stress.
This is the main difference from the simple onion-peeling method described previously. The stress $\sigma_{rz}$ to be reconstructed and the stress $\sigma_{xz}$ to be integrated are linked by an angle $\theta$, as in Eq. (\ref{eq:rotation}) \cite{anton2008}.
This is because photoelasticity can measure only the stress component projected on the plane perpendicular to the camera's optical axis.
In the $N$th ring, $V_2^{(N)}$ is the integral of the shear stress $\sigma_{rz}^{(N)}$ acting on the $N$th ring, i.e., 
\begin{equation}
    V_2^{(N)} = 2C\int^{W_{N,N}}_{-W_{N,N}}\sigma_{xz}^{(N)}dy = 4CW_{N,N}\sigma_{rz}^{(N)} \cos \theta_{N,N},
\end{equation}
where $W_{N,N}$ is the light path of the $N$th camera's optical axis passing through the $N$th ring.
From this equation, the shear stress $\sigma_{rz}^{(N)}$ acting on the $N$th ring can be determined.
The shear stress of the $N$-$1$th ring $\sigma_{rz}^{(N-1)}$, using the previously determined shear stress of the $N$th ring $\sigma_{rz}^{(N)}$, can be determined from the following equation:
\begin{equation}
\begin{split}
    V_2^{(N-1)} = 4CW_{N-1,N-1}\sigma_{rz}^{(N-1)} \cos \theta_{N-1,N-1} \\
    + 4CW_{N-1,N}\sigma_{rz}^{(N)}\cos\theta_{N-1,N}
\end{split}
\end{equation}
In the same way, the shear stress of the $i$th ring can be expressed as
\begin{equation}
    V_2^{(i)} = 4C\sum^{N}_{j=i}W_{i,j}\sigma_{rz}^{(j)}\cos\theta_{i,j},  \label{eq:sum_o_rz}
\end{equation}
where
\begin{equation}
  \theta_{i,j}=
  \begin{cases}
    0 & j=i, \\
    \cos^{-1}\frac{2i}{2j-1}  & j>i.
  \end{cases}
\end{equation}
The axial stress $\sigma_{zz}$ can be determined similarly.
From Eq. (\ref{eq:tomo2}), the axial stress acting on the $i$th ring $\sigma_{zz}^{(i)}$ can be determined using the following equation:
\begin{align}
    V_1^{(i)} - \frac{1}{2\Delta z} \sum^{i}_{j=1} ( V'^{(j)}_2 - V^{(j)}_2 ) \Delta x = 2C\sum^{i}_{j=1} W_{i,j} \sigma_{zz}^{(j)} \label{eq:sum_o_zz}
\end{align}
In the literature \cite{anton2008}, it has been shown that the stresses can be calculated recursively from the outermost ring inward using Eqs. (\ref{eq:sum_o_rz}) and (\ref{eq:sum_o_zz}), and finally, the shear and axial stresses for all rings can be calculated.

Here, we transform the summation equations, Eqs. (\ref{eq:sum_o_rz}) and (\ref{eq:sum_o_zz}), into a matrix form represented by the product of the stress component vector, the coefficient matrix, and the integrated value vector \cite{ichihara2022}.
Equation (\ref{eq:sum_o_rz}) becomes
\begin{align}
    \boldsymbol{V_2}
    =
    4C
    \boldsymbol{\alpha_{rz}}
    \boldsymbol{\sigma_{rz}},
\end{align}
where
\begin{align}
    \boldsymbol{V_2}
    =
    \begin{bmatrix}
        V_2^{(1)} \\
        V_2^{(2)} \\
        \vdots \\
        V_2^{(N-1)} \\
        V_2^{(N)}
    \end{bmatrix}
    , \quad
    \boldsymbol{\sigma_{rz}}
    =
    \begin{bmatrix}
    \sigma_{rz}^{(1)} \\
    \sigma_{rz}^{(2)} \\
    \vdots \\
    \sigma_{rz}^{(N-1)} \\
    \sigma_{rz}^{(N)}
    \end{bmatrix}.
\end{align}
The coefficient matrix for the shear stress $\boldsymbol{\alpha_{rz}}$ is described as
\begin{align}\label{eq:alpha_rz}
    &\boldsymbol{\alpha_{rz}}
    =
    \begin{bmatrix}
    \alpha_{1,1} & \alpha_{1,2} & \dots & \alpha_{1,N-1} & \alpha_{1,N} \\
    0 & \alpha_{2,2} & \dots & \alpha_{2,N-1} & \alpha_{2,N} \\
    \vdots &  \vdots & \ddots &  \vdots &  \vdots \\
    0 & 0 & \dots & \alpha_{N-1,N-1} & \alpha_{N-1,N} \\
    0 & 0 & \dots & 0 & \alpha_{N,N} 
    \end{bmatrix}, \\
    &{\rm where} \quad \alpha_{i,j} = W_{i,j} \cos \theta_{i,j}. \nonumber
\end{align}
In the same way, the summation equation for the axial stress, Eq. (\ref{eq:sum_o_zz}), can be rewritten  as
\begin{align}
    \boldsymbol{V_1} - \frac{\Delta x}{2\Delta z}\boldsymbol{\beta_{V_2}} = 2C\boldsymbol{\alpha_{zz}\sigma_{zz}},
\end{align}
where
\begin{align}
    \boldsymbol{V_1}
    =
    \begin{bmatrix}
        V_1^{(1)} \\
        V_1^{(2)} \\
        \vdots \\
        V_1^{(N-1)} \\
        V_1^{(N)}
    \end{bmatrix}
    , \quad
    \boldsymbol{\sigma_{zz}}
    =
    \begin{bmatrix}
        \sigma_{zz}^{(1)} \\
        \sigma_{zz}^{(2)} \\
        \vdots \\
        \sigma_{zz}^{(N-1)} \\
        \sigma_{zz}^{(N)} \\
    \end{bmatrix}
\end{align}

\begin{align}
    \boldsymbol{\alpha_{zz}}
    =
    \begin{bmatrix}
        W_{1,1} & W_{1,2} & \dots & W_{1,N-1} & W_{1,N} \\
        0 & W_{2,2} & \dots & W_{2,N-1} & W_{2,N} \\
        \vdots &  \vdots & \ddots &  \vdots &  \vdots \\
        0 & 0 & \dots & W_{N-1,N-1} & W_{N-1,N} \\
        0 & 0 & \dots & 0 & W_{N,N}
    \end{bmatrix}
\end{align}

\begin{align} \label{eq:beta_v2}
    \boldsymbol{\beta_{V_2}}
    =
    \begin{bmatrix}
        1 & 1 & \dots & 1 & 1 \\
        0 & 1 & \dots & 1 & 1 \\
        \vdots &  \vdots & \ddots &  \vdots &  \vdots \\
        0 & 0 & \dots & 1 & 1 \\
        0 & 0 & \dots & 0 & 1
    \end{bmatrix}
    \begin{bmatrix}
        V_2'^{(1)} - V_2^{(1)} \\
        V_2'^{(2)} - V_2^{(2)} \\
        \vdots \\
        V_2'^{(N-1)} - V_2^{(N-1)} \\
        V_2'^{(N)} - V_2^{(N)}
    \end{bmatrix}
\end{align}

The vector of the shear and axial stresses $\boldsymbol{\sigma_{rz}}, \boldsymbol{\sigma_{zz}}$ can be obtained instantaneously by multiplying the vector of the integrated value by the inverse of the coefficient matrices, i.e.,
\begin{align}
    \boldsymbol{\sigma_{rz}} &= \frac{1}{4C} \boldsymbol{ \alpha_{rz}^{-1} V_2 }, \\
    \boldsymbol{\sigma_{zz}} &= \frac{1}{2C}\boldsymbol{\alpha_{zz}^{-1}}\left( \frac{\Delta x}{2\Delta z}\boldsymbol{\beta_{V_2}} - \boldsymbol{V_1} \right)
\end{align}
if the inverse matrices, $\boldsymbol{ \alpha_{rz}^{-1}}$ and $\boldsymbol{\alpha_{zz}^{-1}}$, are calculated in advance.
This reduces the computational time required to calculate stresses acting on all rings compared with the recursive calculation of Eqs. (\ref{eq:sum_o_rz}) and (\ref{eq:sum_o_zz}) from the outermost ring inward.

\subsubsection{Determination of the radial and hoop stresses}

The radial $\sigma_{rr}$ and hoop $\sigma_{\theta\theta}$ stresses are determined using the equations of the theory of linear elasticity \cite{errapart2011,aben2010a}.
The stresses must satisfy the equilibrium equation and compatibility equation,
\begin{align}
    &\frac{\partial \sigma_{rr}}{\partial r} + \frac{\sigma_{rr}-\sigma_{\theta\theta}}{r} + \frac{\partial \sigma_{rz}}{\partial z}= 0, \\
    &\frac{\partial}{\partial r}\{ \sigma_{\theta\theta} - \nu (\sigma_{rr} + \sigma_{zz})\} - (1+\nu)\frac{\sigma_{rr}-\sigma_{\theta\theta}}{r} = 0,
\end{align}
respectively.
The differential equations for the radial and hoop stresses are obtained by transforming these equations as follows:
\begin{align}
    &\frac{\partial \sigma_{rr}}{\partial r} = - \frac{\sigma_{rr}-\sigma_{\theta\theta}}{r} - \frac{\partial \sigma_{rz}}{\partial z}, \label{eq:diff_rr}\\
    &\frac{\partial \sigma_{\theta\theta}}{\partial r} =  \frac{\sigma_{rr}-\sigma_{\theta\theta}}{r} - \nu\frac{\partial \sigma_{rz}}{\partial z} + \nu\frac{\partial \sigma_{zz}}{\partial r}. \label{eq:diff_tt}
\end{align}
Since the spatial distribution of the shear and axial stresses was obtained in Sec. \ref{sec:Determination_o_rz_o_zz}, the second and subsequent terms on the right-hand side of these equations are known values.
Therefore, by solving this differential equation numerically, the radial and hoop stresses can be determined.
Discretizing Eqs. (\ref{eq:diff_rr}) and (\ref{eq:diff_tt}) using the first-order forward-difference method yields
\begin{align}
    \sigma_{rr}^{(i)} = \sigma_{rr}^{(i-1)} + \Delta r & \left\{- \frac{\sigma_{rr}^{(i-1)}}{r^{(i-1)}} + \frac{\sigma_{\theta\theta}^{(i-1)}}{r^{(i-1)}}\right. \nonumber \\
    &\quad \left. - \left(\frac{\partial \sigma_{rz}}{\partial z}\right)^{(i-1)} \right\}, \label{eq:disc_rr}
\end{align}
\begin{align}
    \sigma_{\theta\theta}^{(i)} =& \sigma_{\theta\theta}^{(i-1)} + \Delta r \left\{\frac{\sigma_{rr}^{(i-1)}}{r^{(i-1)}} + \frac{\sigma_{\theta\theta}^{(i-1)}}{r^{(i-1)}}\right. \nonumber \\
    & \left. - \nu \left(\frac{\partial \sigma_{rz}}{\partial z}\right)^{(i-1)} \right\} + \nu (\sigma_{zz}^{(i)} - \sigma_{zz}^{(i-1)}), \label{eq:disc_tt}
\end{align}
where $i = 1$ for $r = 0$ and $i$ increases toward $r$.
The radial and hoop stresses can be determined by setting appropriate boundary conditions.

\subsubsection{Quasi-static approximation}\label{sec:quasi-static approximation}

The tomography algorithm described in the previous sections assumed equilibrium except for the case of the reconstruction of the shear stress $\sigma_{rz}$.
Therefore, the quasi-static approximation has to be adopted to reconstruct stresses from time-series data of the measured photoelastic parameters.
The impact velocity $V$ and the pressure wave speed $v_p$ in the material are compared to confirm this approximation.
The pressure wave speed $v_p$ can be expressed using the following equation \cite{johnson1985}:
\begin{equation}\label{eq:quasi-static}
    v_p = \sqrt{E/\rho_s},
\end{equation}
where $E$ and $\rho_s$ are Young's modulus and the density of the substrate material, respectively.
For low-velocity impacts, where the sphere velocity $V$ is much smaller than the pressure wave velocity $v_p$, the quasi-static approximation is more likely to be valid.

\subsection{Experiment}

\begin{figure*}[t]
    \centering
    \includegraphics[width=0.8\textwidth]{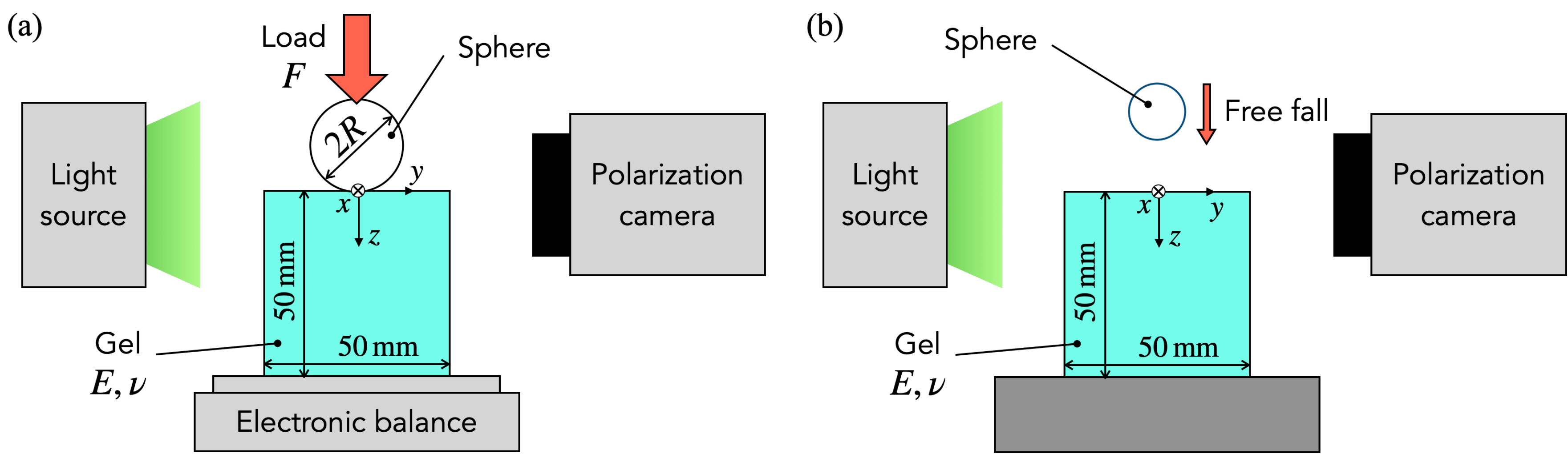}
    \caption{A schematic of the experimental setup for (a) static Hertzian contact and (b) dynamic Hertzian contact.}
    \label{fig:setup}
\end{figure*}

\begin{figure}[t]
    \centering
    \includegraphics[width=0.8\columnwidth]{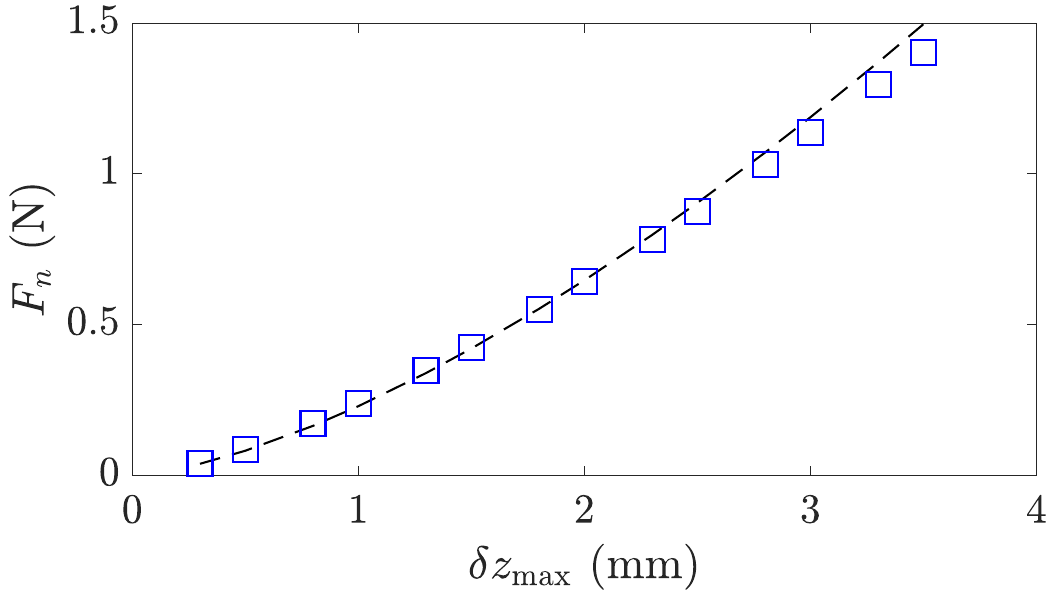}
    \caption{The relationship between the surface deformation $\delta z_{\rm max}$ of the gel and the loading normal force $F_n$ when the sphere, which has a radius $R$ of $7.35$ mm, is pressed against it. The dashed line indicates the theoretical normal force from Eq. (\ref{eq:SurfaceDeformation}) using the measured surface deformation $\delta z_{\rm max}$ and the Young's modulus of $E = 47.4$ kPa.}
    \label{fig:SurfaceDeformation}
\end{figure}

The schematics of the setups for the experimental verification in the static and dynamic cases are shown in Figs. \ref{fig:setup}(a) and \ref{fig:setup}(b), respectively.
The setup includes a light source, a measurement target, and a polarization camera.
A urethane gel block (urethane gel phantom, Exseal Co., Ltd., Gifu, Japan), which has a size of $50 \times 50 \times 50$ $\rm mm^3$ and a density of $\rho_s$ of $1064$ $\rm kg/m^3$, is placed on an electronic balance.
We estimated the Young's modulus of the gel, $E$, from the surface deformation $\delta z_{\rm max}$ along the $z$-axis and the loading normal force $F_n$ with the electric balance using Hertzian contact theory \cite{johnson1985}, i.e.,
\begin{equation}\label{eq:SurfaceDeformation}
    \delta z_{\rm max}
    =
    \left( \frac{9F_n^2}{16 {E^{\ast}}^2 R} \right)^{1/3},
\end{equation}
where $E^\ast=E/(1-\nu^2)$.
The Young's modulus of the gel was determined to be $47.4 \, \rm kPa$, with the Poisson's ratio of the gel $\nu$ assumed to be 0.5 in this study.
Note that Hertzian contact theory assumes that the contact radius (the area where the sphere touches the substrate) is considerably smaller than $R$ \cite{popov2017, johnson1985}.
However, the measured surface deformation results closely follow the curve outlined in Eq. (\ref{eq:SurfaceDeformation}) (see Fig. \ref{fig:SurfaceDeformation}).
Our previous paper delves into the Hertzian contact problem in the context of a highly deformable substrate \cite{mitchell2022}.

For the static case (Fig.\ref{fig:setup}(a)), a styrol sphere with a diameter of $2R = 14.7 \, \rm mm$ is vertically pressed against the surface of the gelatin with a loading normal force $F_n$.
This experimental setup produces Hertzian contact between a sphere and a half-space \cite{johnson1985}. 
The loading normal force $F_n$ is calculated as $F_n=mg$, where the apparent mass $m$ is measured using the electronic balance.
The acceleration of gravity used in the calculation is $9.81 \, \rm m/s^2$.
The retardation and orientation fields under sphere pressing are recorded by the polarization camera (Photron, CRYSTA PI-5WP) with the light source (Thorlabs, SOLIS-565C).
To reduce the deterioration of the measurement accuracy due to the bandwidth of the light source wavelength \cite{su2023a}, a narrow bandwidth bandpass filter (Photron, with a half-bandwidth of approximately 20 nm) is installed in front of the camera sensors.
Then, the typical wavelength of the light is 540 nm.
The number of pixels in the measured images of the photoelastic parameters is $191 \times 208$ pixels in the region of $1.37R \times 1.50R$, which is used for stress reconstruction.
The spatial resolution of the photoelastic parameters images is $52.7$ \textmu m/pix for the static case.

For the dynamic case (Fig.\ref{fig:setup}(b)), a plastic sphere with a radius of $R = 2.98 \, \rm mm$ and a density of $\rho = 997.3 \, \rm kg/m^3$ impacts the gel during free fall.
The impact velocity of the sphere, $V$, is varied from 0.3 to 0.7 m/s by adjusting the falling height of the sphere.
The retardation and orientation fields during the sphere impact are recorded by the polarization camera (Photron, CRYSTA PI-1P) with the light source (Thorlabs, SOLIS-525C).
The typical wavelength of the light is set to 520 nm by using a bandpass filter placed in front of the lens attached to the camera.
The temporal resolution of the camera is 20,000 f.p.s for high-speed stress measurement.
The number of pixels in the measured images of the photoelastic parameters is $115 \times 171$ pixels in the region of $3.28R \times 5.14R$, which is used for stress reconstruction.
The spatial resolution of the photoelastic parameter images is $84.9$ \textmu m/pix for the dynamic case.

\begin{figure*}[t]
    \centering
    \includegraphics[width=0.9\textwidth]{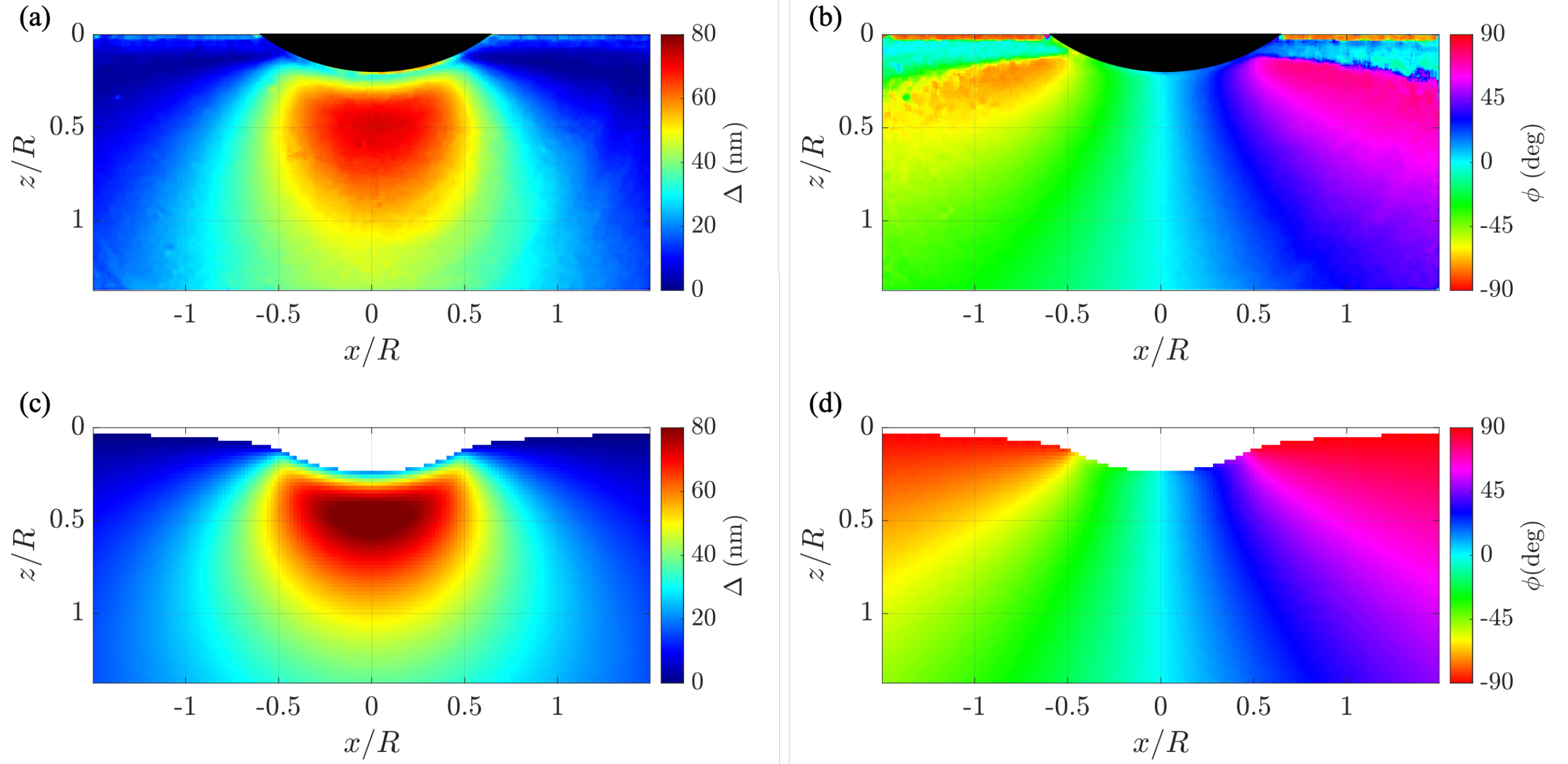}
    \caption{An example of the measured (a) retardation and (b) orientation fields. The loading normal force $F_n$ of 0.55 $\rm N$ is induced by the pressing of a sphere with radius $R= 7.35$ mm. The black parts in the upper middle area indicate the pressed sphere. (c) and (d) are the theoretical retardation and orientation fields, which can be obtained by integrating the theoretical stress field calculated by solving the Hertzian contact problem with the stress-optic coefficient $C = 1.14 \times 10^{-9} \, \rm Pa^{-1}$. The white parts in the upper area indicate the deformed surface of the gel due to sphere pressing.}
    \label{fig:measured_image}
\end{figure*}

The light source generates incident light of a typical wavelength $\lambda$.
After passing through the polarizer, which has an angle of $0^\circ$, the incident light then goes through a quarter-wave plate, which has an angle of $45^\circ$, resulting in circular polarization.
The stressed gel allows circularly polarized light to pass through and emits it as elliptically polarized light with retardation $\Delta$ and orientation $\phi$.
Using a polarization camera, it is possible to simultaneously measure both the orientation and retardation of the emitted light.
\begin{figure}[t]
    \centering
    \includegraphics[width=1\columnwidth]{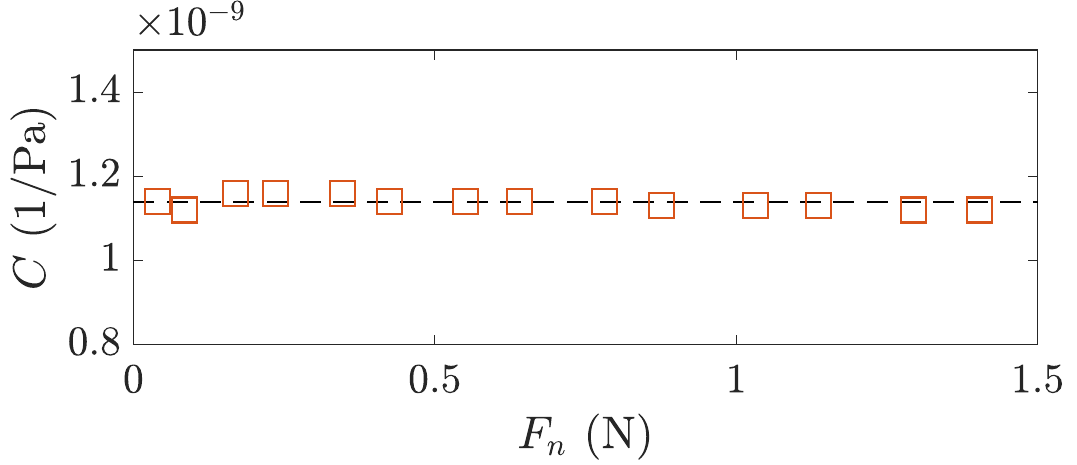}
    \caption{The stress-optic coefficient $C$ with different loading normal forces $F_n$. The dashed line indicates the mean value of the stress-optic coefficient of the urethane gel used in this study, which is $1.14 \times 10^{-9} \, \rm Pa^{-1}$.}
    \label{fig:StressOpticCoefficient}
\end{figure}
Four linear polarizers with angles of $I_{0^\circ}, I_{45^\circ}, I_{90^\circ}$, and $I_{135^\circ}$ are installed in the four neighboring pixels of the image sensor of the polarization camera.
This set of four neighboring pixels is called a ``super-pixel'' \cite{lane2022a}.
The intensity values measured by the camera's sensor through each linear polarizer are denoted by $I_{0^\circ}, I_{45^\circ}, I_{90^\circ}$, and $I_{135^\circ}$, respectively.
Using the four-step phase-shifting method \cite{ramesh2021,otani1994,onuma2014,yokoyama2023}, the retardation $\Delta$ and orientation $\phi$ are obtained from the four intensity values of the super-pixel as follows:
\begin{gather}
\Delta = \frac{\lambda}{2\pi} \sin^{-1}{\frac{\sqrt{\left(I_{90^\circ}-I_{0^\circ}\right)^2+\left(I_{45^\circ}-I_{135^\circ}\right)^2}}{I/2}}, \label{eq:delta} \\
\phi = \frac{1}{2}\tan^{-1}\frac{I_{90^\circ}-I_{0^\circ}}{I_{45^\circ}-I_{135^\circ}}, \label{eq:phi}
\end{gather}
where $I = I_{0^\circ} + I_{45^\circ} + I_{90^\circ} + I_{135^\circ}$.
As an example of the measured data, the retardation and orientation are calculated using software (Photron Ltd., CRYSTA Stress Viewer) and are shown in Figs. \ref{fig:measured_image}(a) and \ref{fig:measured_image}(b).
The retardation and orientation data are filtered to reduce the noise using Matlab's median filter function (medfilt2).
The details of the principle of measurement are also described in Refs. \cite{onuma2012,onuma2014,yokoyama2023}.

The theoretical retardation and orientation field can be obtained using the integrated photoelasticity \cite{yokoyama2023} from the theoretical solution of the stress fields calculated using the Hertzian contact problem \cite{johnson1985,ike2019,mitchell2023,yokoyama2023}, for which an example of the theoretical data is shown in Figs. \ref{fig:measured_image}(c) and \ref{fig:measured_image}(d). The retardation and orientation fields are calculated using the same conditions as in the experiment shown in Figs. \ref{fig:measured_image}(a) and \ref{fig:measured_image}(b) using a stress-optic coefficient of $1.14 \times 10^{-9} \, \rm Pa^{-1}$.
The stress-optic coefficient was determined by comparing the experimental and theoretical retardation fields, and chosen to be the value that minimizes the root mean square error between the calculated and experimental retardation.
The procedure for determining $C$ is described in our previous paper \cite{yokoyama2023}.
The result of the determination of the stress-optic coefficient is shown in Fig. \ref{fig:StressOpticCoefficient}.
The stress-optic coefficient is almost constant over a wide range of loading normal forces $F_n$.

\section{Results and Discussion}\label{sec:ResultandDiscussion}

\begin{figure*}[t]
    \centering
    \includegraphics[width=0.9\textwidth]{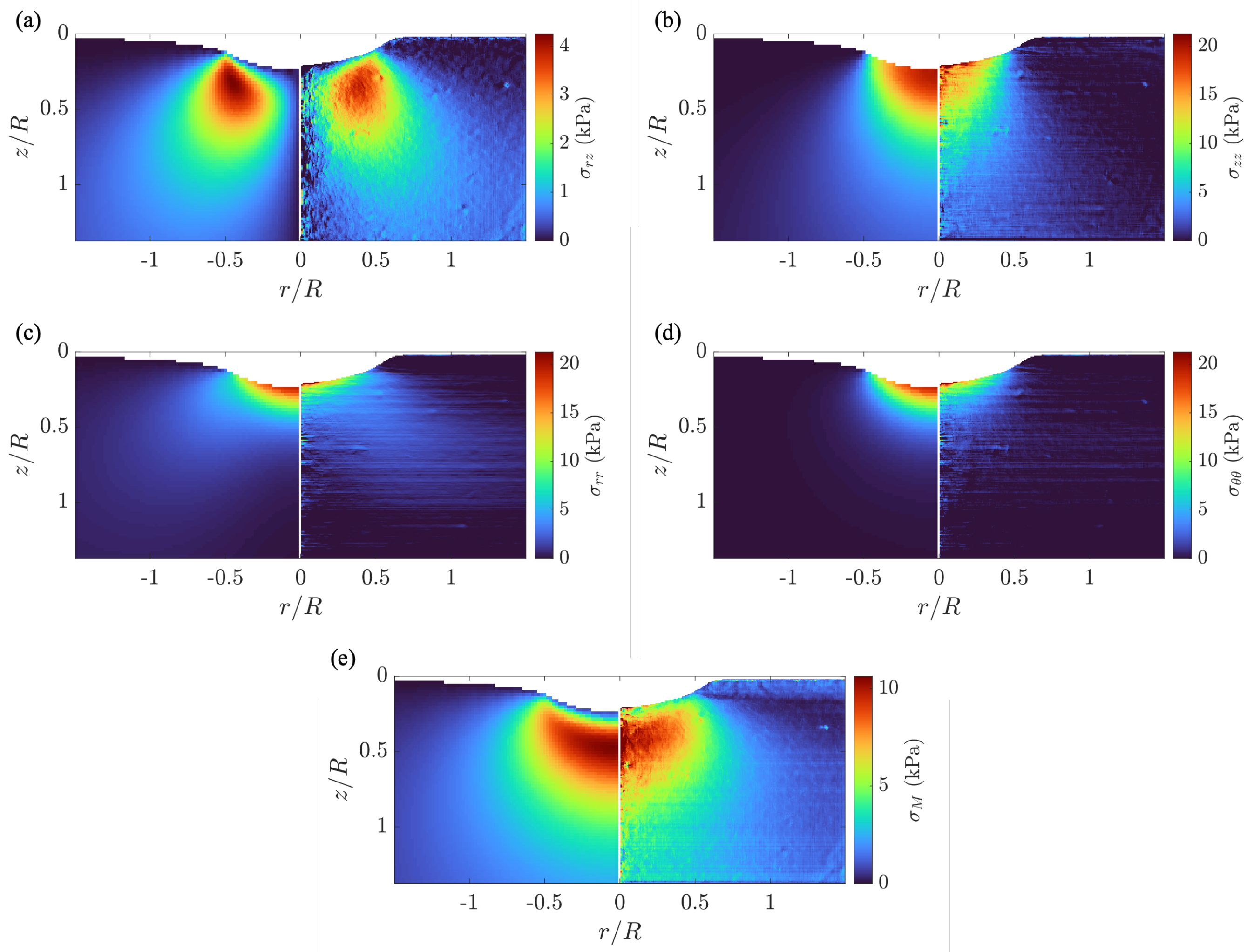}
    \caption{A comparison between the (left-hand side) theoretical and (right-hand side) experimental stress fields. (a), (b), (c), and (d) show the shear, axial, radial, and hoop stresses, respectively. (e) The von Mises stress calculated from the stress components via Eq. (\ref{eq:vonMises}). The loading normal force $F_n$ is $0.55 \, \rm N$, induced by pressing the of a sphere with radius $R= 7.35$ mm. The white parts in the upper area indicate the deformed surface of the gel due to sphere pressing.}
    \label{fig:reconstructed_fields}
\end{figure*}

\subsection{Experimental validation of photoelastic tomography for static Hertzian contact}

In this section, we demonstrate that photoelastic tomography can reconstruct all the components of the axisymmetric stress fields, $\sigma_{rr}$, $\sigma_{\theta\theta}$, $\sigma_{zz}$, and $\sigma_{rz}$, in a soft material that shows large deformation, focusing on the static Hertzian contact problem where a rigid sphere is pressed against an elastic half-space.

Figure \ref{fig:reconstructed_fields} compares the reconstructed stress fields (right-hand side), which are from the experimental photoelastic parameter data, and the theoretical stress fields (left-hand side), which are obtained by solving the Hertzian contact problem using similar conditions to the experiment for $m$, $R$, and $E$.
For all stress components, there are errors near the surface of the gel, but overall, the experimental and theoretical results agree well.
It should be noted that the experimental results do not necessarily completely agree with the theoretical results because the contact conditions between the sphere and the gel surface are not necessarily identical in the theoretical calculations and the experiments.
In other words, errors between the experimental and theoretical results are expected to be a result of the surface contact conditions in the experiment, which are not considered in the calculation of Hertzian contact theory.
The following issues can be considered as the sources of the noisy distribution of the shear stress $\sigma_{rz}$ around the $z$-axis.
The orientation values approach zero around the $z$-axis (see Fig. \ref{fig:measured_image}(b)) and are sometimes negative.
Theoretically, the azimuth should be greater than or equal to zero on the right-hand side of the $z$-axis.
Therefore, the filtering of the orientation data before reconstruction and the determination of the $z$-axis position has a significant influence on this noise.
Additionally, the axial stress $\sigma_{zz}$ distribution shows horizontal stripe noise, while the shear stress $\sigma_{rz}$ does not show any.
This is because the $z$-axis gradient of the $V_2$ field, i.e., $V_2'-V_2$, is used when reconstructing the axial stress (see Eq. (\ref{eq:sum_o_zz})).
Since the $V_2$ distribution is not perfectly smooth, the $V_2'-V_2$ distribution is discontinuous along the $z$-axis.
Therefore, the axial stress calculated using this distribution is also not smooth in the $z$-direction, resulting in horizontal stripe noise.
This noise carries over to the radial and hoop stresses.

\begin{figure*}[t]
    \centering
    \includegraphics[width=1\textwidth]{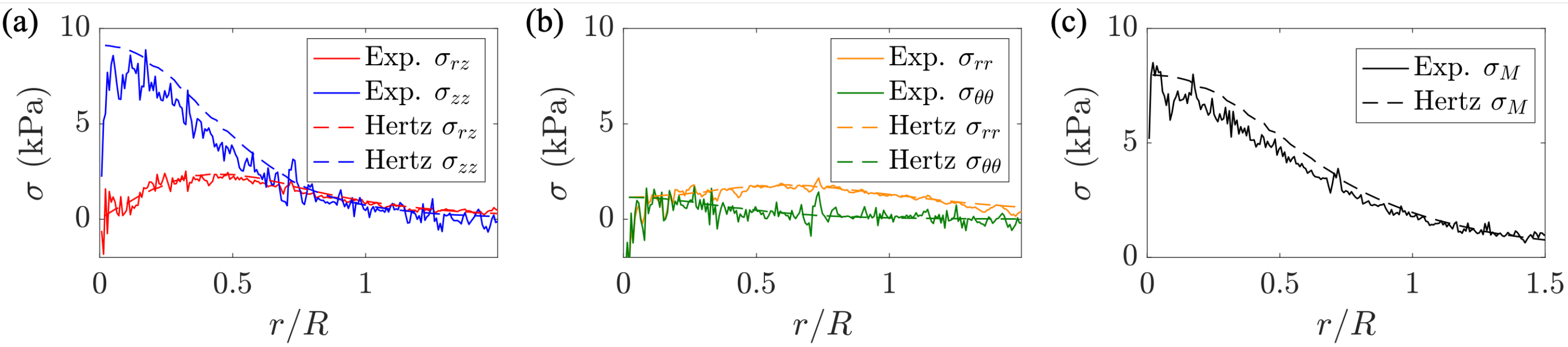}
    \caption{The line profiles of the reconstructed stress field compared with that of the expected stress field at $z/R = 0.7$. (a) The shear and axial stresses. (b) The radial and hoop stresses. (c) The von Mises stress. The loading normal force $F_n$ is $0.55 \, \rm N$, and is induced by the pressing of a sphere with radius $R= 7.35$ mm.}
    \label{fig:reconstructed_lineprofiles}
\end{figure*}

Since all stress components have now been determined, the von Mises stress $\sigma_{M}$, which is often used as a failure criterion for ductile materials and as a force criterion related to plastic deformation \cite{mitchell2023}, can also be determined using the equation
\begin{equation}\label{eq:vonMises}
    \sigma_{M} = \sqrt{3J_2},
\end{equation}
where
\begin{equation}
\begin{aligned}
    J_2 = \frac{1}{6}  & \left[ (\sigma_{rr} - \sigma_{\theta\theta})^2 + (\sigma_{\theta\theta}-\sigma_{zz})^2+(\sigma_{zz}-\sigma_{rr})^2\right]\\
    & + \sigma_{r\theta}^2+\sigma_{\theta z}^2+\sigma_{rz}^2.
\end{aligned}
\end{equation}
Figure \ref{fig:reconstructed_fields}(e) compares the theoretical and experimental von Mises stresses and shows good agreement.

Figure \ref{fig:reconstructed_lineprofiles} shows the line profiles of the reconstructed stress fields compared with the theoretical solutions at a certain $z$ position, $z/R = 0.7$.
The reconstructed stress fields show similar distributions to those estimated by solving the Hertzian contact problem.
The noisy distribution of the line profiles is mainly due to the noise of the photoelastic parameters (the retardation and orientation fields).
In particular, in the calculation of the axial stress $\sigma_{zz}$, the integral value vector expressed in Eq. (\ref{eq:beta_v2}) is sensitive to the noise of the photoelastic parameters because it is the amount of change between adjacent pixels in the $z$-direction of $V_2$ consisting of photoelastic parameters.
Smoothing the experimental data with a more appropriate filter is expected to reduce this noise.

These results indicate that photoelastic tomography can reconstruct all components of axisymmetric stress fields in soft materials with large deformation.
This allows for the estimation of the von Mises stresses, which are important for evaluating the yield value of a material, and the analysis of principal stresses and their directions (rather than the secondary principal stress).

\subsection{High-speed stress field measurement during dynamic Hertzian contact}\label{sec:RD_dynamic}

\begin{figure*}[t]
    \centering
    \includegraphics[width=0.9\textwidth]{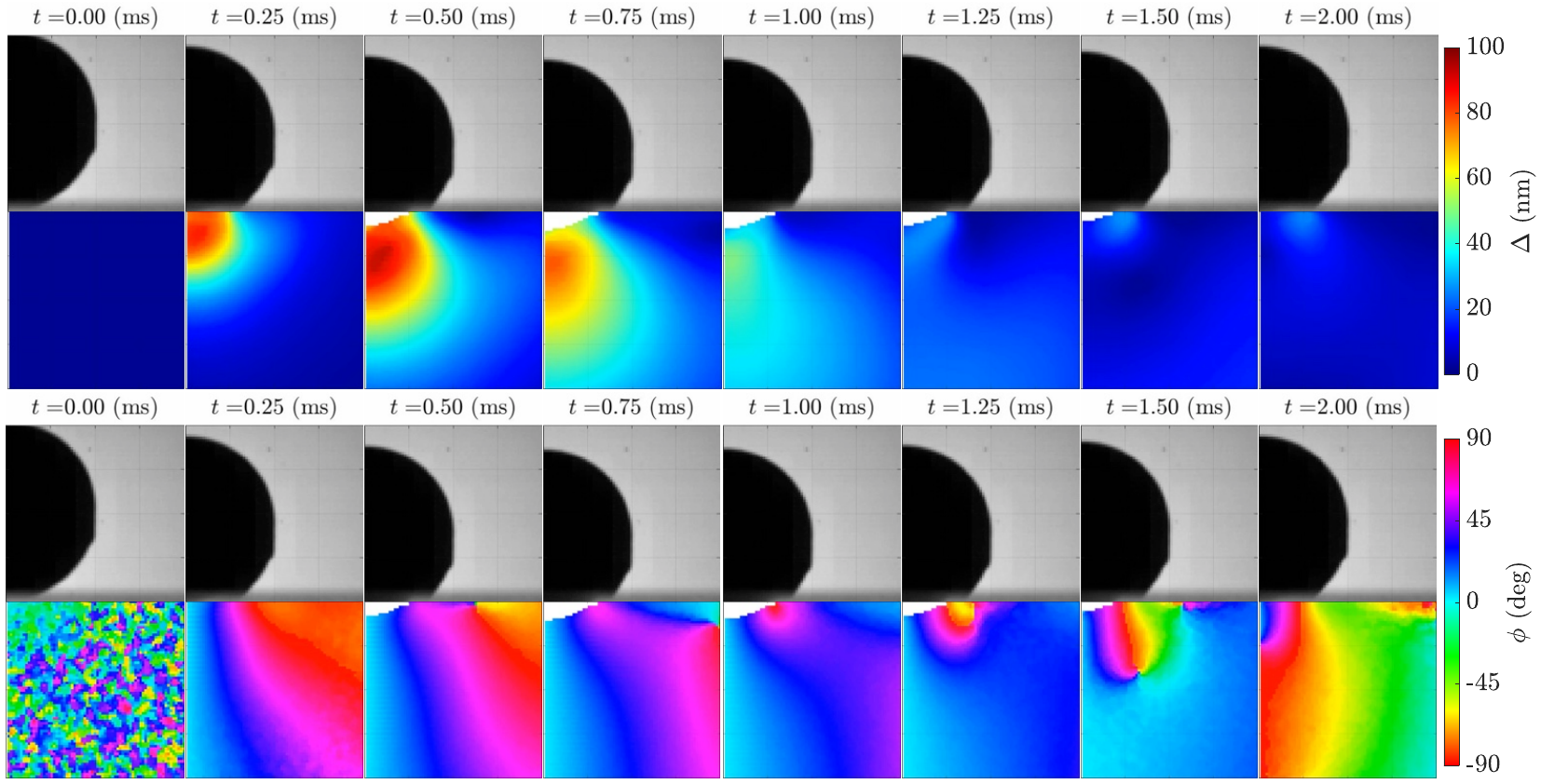}
    \caption{The temporal evolution of the retardation and orientation fields during the impact of a plastic sphere. The impact velocity of a sphere with radius $R= 2.98$ mm is $V=2.2 \, \rm m/s$. The white parts in the middle area indicate the deformed surface of the gel due to the sphere impact.}
    \label{fig:RetardationOrientationFields_SphereImpact}
\end{figure*}

In this section, we determine the applicability of photoelastic tomography for high-speed stress field measurements in soft materials, focusing on the dynamic Hertzian contact problem, in which a rigid sphere impacts an elastic half-space.

Firstly, the validity of the quasi-static approximation in this dynamic case was checked by comparing the sphere's impact velocity $V$ and the substrate's pressure wave speed $v_p$ (Sec. \ref{sec:quasi-static approximation}).
In our experiment, the maximum value of $V/v_p$ is about 0.4 because the maximum $V$ is 2.7 m/s while $v_p$ is approximately 6.7 m/s, since $E = 47.4$ kPa and $\rho_s = 1064 \, \rm kg/m^3$.
Therefore, we assumed that the quasi-static approximation holds in our experiment since the ratio $V/v_p$ is smaller than 1.

Figure \ref{fig:RetardationOrientationFields_SphereImpact} shows the retardation and orientation distribution in a gel when a plastic sphere with a radius $R = 2.9$ mm makes impact at a velocity $V \simeq 2.2$ m/s.
The retardation and orientation corresponding to the dynamic stress field in the soft material caused by the sphere impact are measured.
The white area indicates the gel's surface deformation due to the sphere's penetration.
Before impact ($t=0$ ms), because there is no stress in the gel, the retardation is zero, and the orientation is random from $-90^\circ$ to $90^\circ$.
The maximum retardation can be observed at around $t = 0.5 \sim 0.75$ ms, when the vertical displacement of the sphere reaches its maximum value.
Then, the retardation decreases with time according to the upward motion of the sphere.

\begin{figure*}[t]
    \centering
    \includegraphics[width=0.9\textwidth]{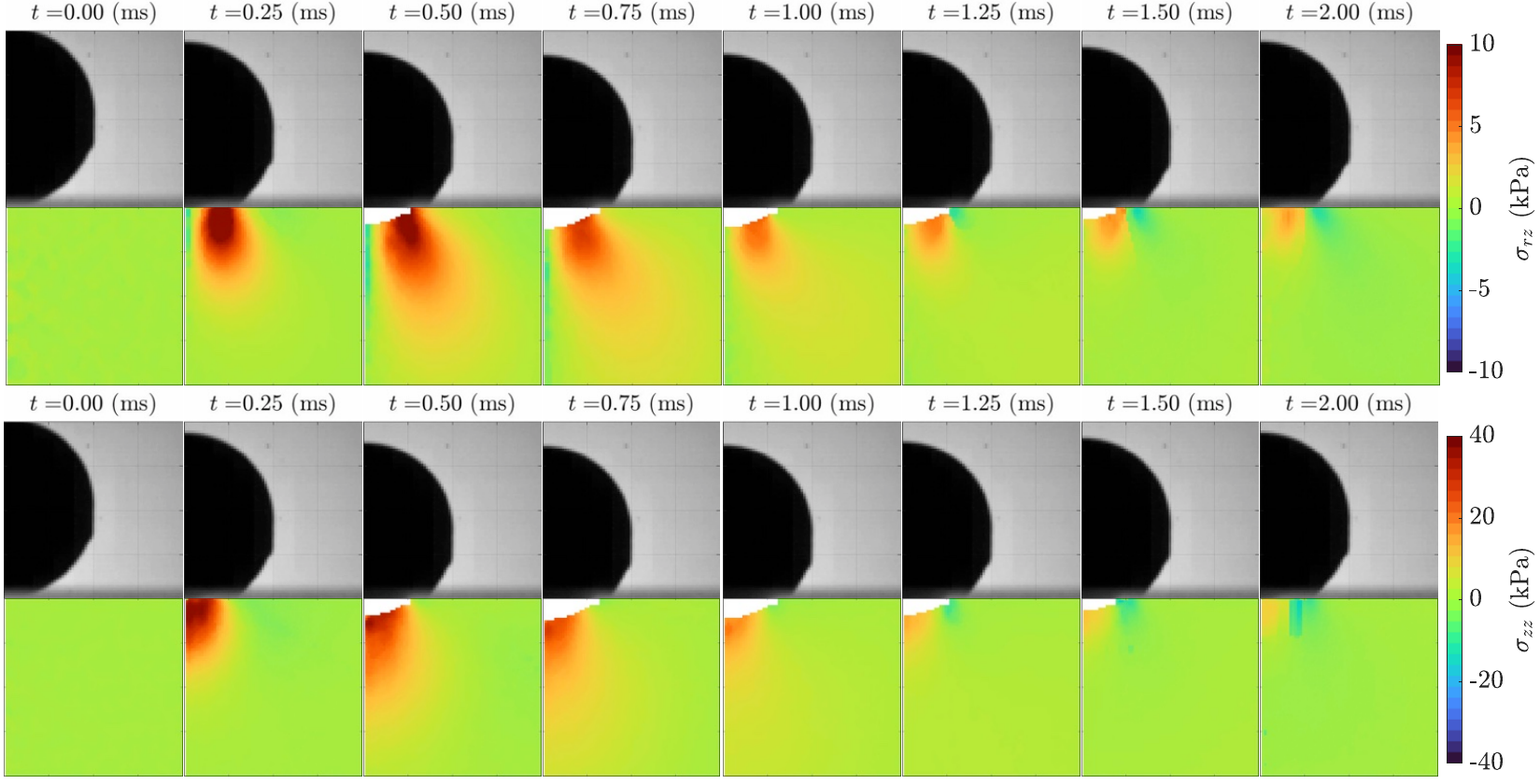}
    \caption{The temporal evolution of the (upper row) shear and (lower row) axial stresses during the impact of a plastic sphere. The impact velocity of the sphere with radius $R= 2.98$ mm is $V=2.2 \, \rm m/s$. The white parts in the middle area indicate the deformed surface of the gel due to the sphere impact.}
    \label{fig:ReconstractedFields_ShearAxial_unsteady}
\end{figure*}

\begin{figure*}[t]
    \centering
    \includegraphics[width=0.9\textwidth]{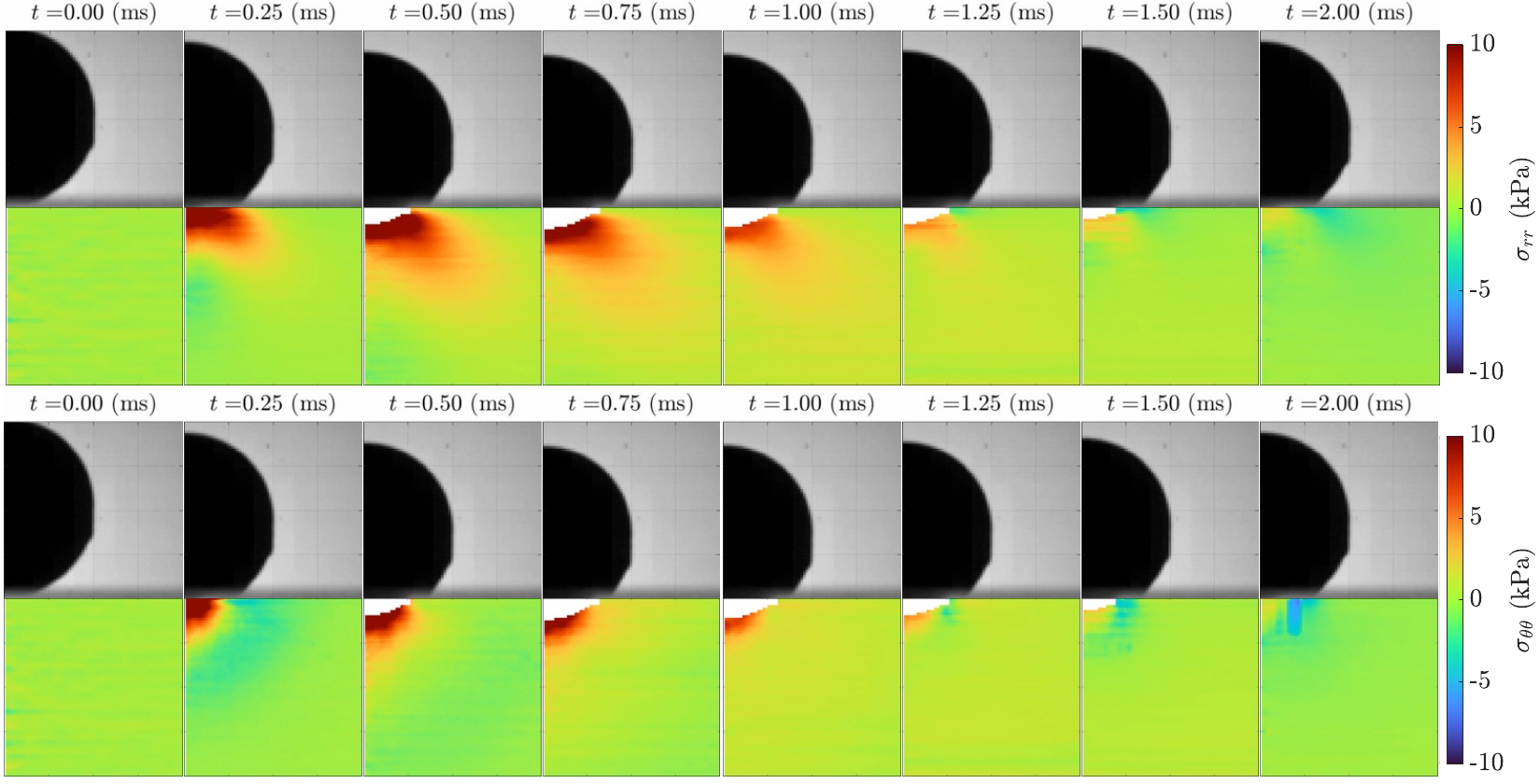}
    \caption{The temporal evolution of the (upper row) radial and (lower row) hoop stresses during the impact of a plastic sphere. The impact velocity of the sphere with radius $R= 2.98$ mm is $V=2.2 \, \rm m/s$. The white parts in the middle area indicate the deformed surface of the gel due to the sphere impact.}
    \label{fig:ReconstractedFields_RadialHoop_unsteady}
\end{figure*}

Figures \ref{fig:ReconstractedFields_ShearAxial_unsteady} and \ref{fig:ReconstractedFields_RadialHoop_unsteady} show the temporal evolution of the shear, axial, radial, and hoop stress distributions during the impact of the sphere, as reconstructed from the photoelastic parameters (Fig. \ref{fig:RetardationOrientationFields_SphereImpact}).
The accuracy of the reconstructed stress fields was validated by comparing the scaling law for the maximum impact force $F_{n, \rm max}$ calculated from the axial stress field with the theoretical prediction.
The details of this comparison are discussed in the next section (Sec. \ref{sec:ScalingLaw}).
A stress maximum appears between 0.25 and 0.5 ms for all stress components.
After that, the stresses decrease with time following the upward motion of the sphere.
For the shear and axial stresses, negative values eventually appear near the edges of the contact area, probably due to the adhesion force between the spheres and the gel surface.
For the radial stress, the region of the acting stress is wider than the others, and a negative stress region appears along the $z$-direction.
The hoop stress exhibits both negative and positive values immediately following the impact, and then the negative stress is damped.

\begin{figure}[t]
    \centering
    \includegraphics[width=1\columnwidth]{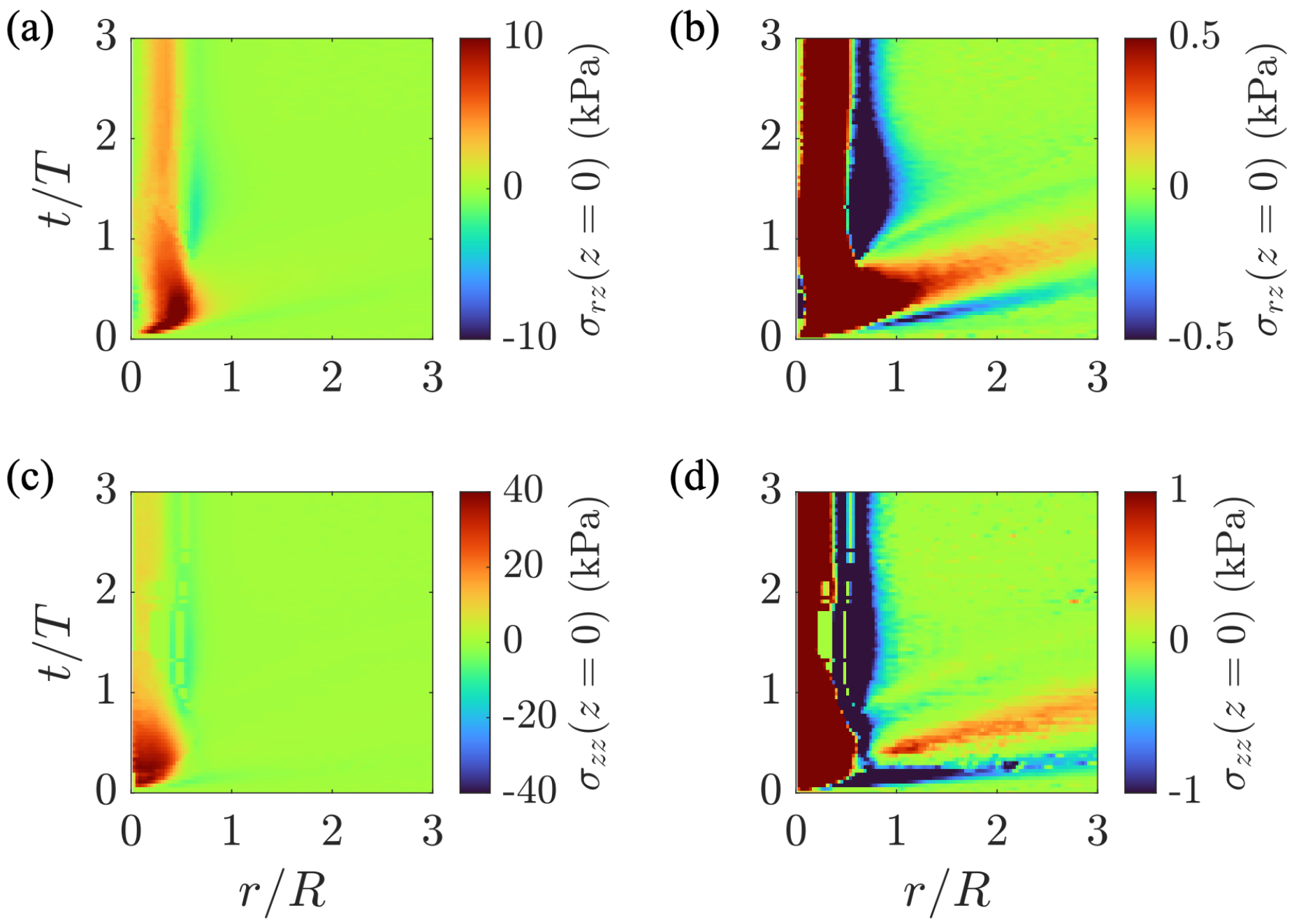}
    \caption{The kymographs of (a,b) the shear and (c,d) the axial stresses acting on the substrate surface, $\sigma_{rz}(r,z=0,t)$ and $\sigma_{zz}(r,z=0,t)$, for a sphere radius $R= 2.98$ mm and impact velocity $V=2.2 \, \rm m/s$, with different color bar limits.}
    \label{fig:surface_stress}
\end{figure}

Figure \ref{fig:surface_stress} shows kymographs of the shear and axial stresses acting on the substrate surface, $\sigma_{rz}(r,z=0,t)$ and $\sigma_{zz}(r,z=0,t)$.
Figures \ref{fig:surface_stress}(b) and \ref{fig:surface_stress}(d) are plotted using the same data as in Figs. \ref{fig:surface_stress}(a) and \ref{fig:surface_stress}(c), respectively, with different color bar limits to visualize stress values with small magnitudes.
The waves of the shear and axial stresses at the surface show different velocities.
This cannot be obtained from the visualization of the retardation and orientation data measured by the polarization camera alone, such as in Fig. \ref{fig:RetardationOrientationFields_SphereImpact}.
It can only be quantified by reconstructing the stress field using these data.

Figures \ref{fig:ReconstractedFields_ShearAxial_unsteady}, \ref{fig:ReconstractedFields_RadialHoop_unsteady}, and \ref{fig:surface_stress} show that even stress fields with a wide range from $\mathcal{O}(10^{-1})$ to $\mathcal{O}(10^1)$ kPa can be measured simultaneously.
The measurable range of stress depends on the wavelength of the light source $\lambda$ and the stress-optic coefficient of the material used, $C$.
From Eq. (\ref{eq:delta}), the measurement range of retardation $\Delta$ is limited to $0-\lambda/4$.
Also, from Eqs. (\ref{eq:V1}) and (\ref{eq:V2}), the magnitude of $\Delta$ is determined by the stress-optic coefficient $C$.
Therefore, the order of magnitude of the measurable stress can be changed by using a different gel, i.e., changing the stress-optic coefficient $C$.
Gelatin gel has an order of magnitude larger $C$ \cite{yokoyama2023,aben1993a,harris1978,bayley1959} than the urethane gel used in this experiment and is more sensitive to small stresses.
Additionally, the range of measurable retardation could be further expanded by using a color polarization camera \cite{zhang2024}.

These results demonstrate that photoelastic tomography can be applied successfully even in fields where the stress tensor components are fast and time-varying, such as stress wave propagation.

\subsection{The scaling law for the maximum impact force}\label{sec:ScalingLaw}

In this section, the normal impact force acting on the gel surface is estimated, and the scaling laws for its maximum values are compared with theoretical predictions from dynamic Hertzian contact to evaluate the accuracy of the photoelastic tomography in reconstructing the time-varying stress fields.

\begin{figure}[t]
    \centering
    \includegraphics[width=1\columnwidth]{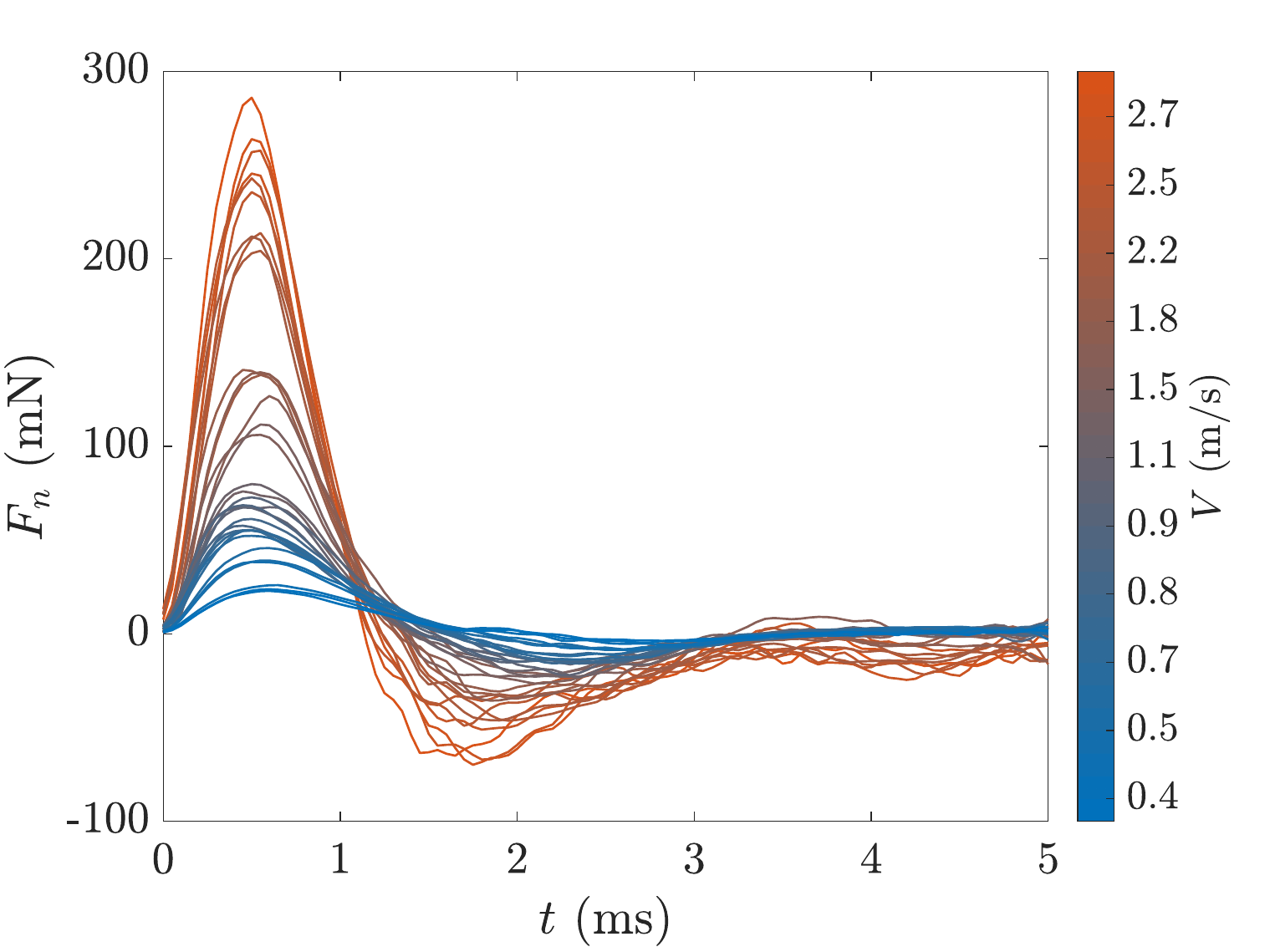}
    \caption{The temporal evolution of the normal force acting on the surface during the impact of a sphere at different impact velocities.}
    \label{fig:ImpactForce}
\end{figure}

The normal impact force $F_n$ can be obtained by integrating over the axial stress $\sigma_{zz}$ acting on the surface of the gel, i.e.,
\begin{equation}
    F_n(t) = 2 \pi \int^\infty_0 \sigma_{zz} r dr.
\end{equation}
Figure \ref{fig:ImpactForce} shows the temporal evolution of the normal impact force $F_n$ for different impact velocities.
The normal impact force suddenly increases after impact and reaches its maximum value.
Then, it decreases following the sphere's vertical motion because of the gel's viscous damping effect.
The maximum normal impact force increases with the impact velocity.
In the case of a high impact velocity, the impact force becomes negative when the sphere moves upward.
This is likely due to the adhesion force between spheres and the gel surface.

\begin{figure}[t]
    \centering
    \includegraphics[width=1\columnwidth]{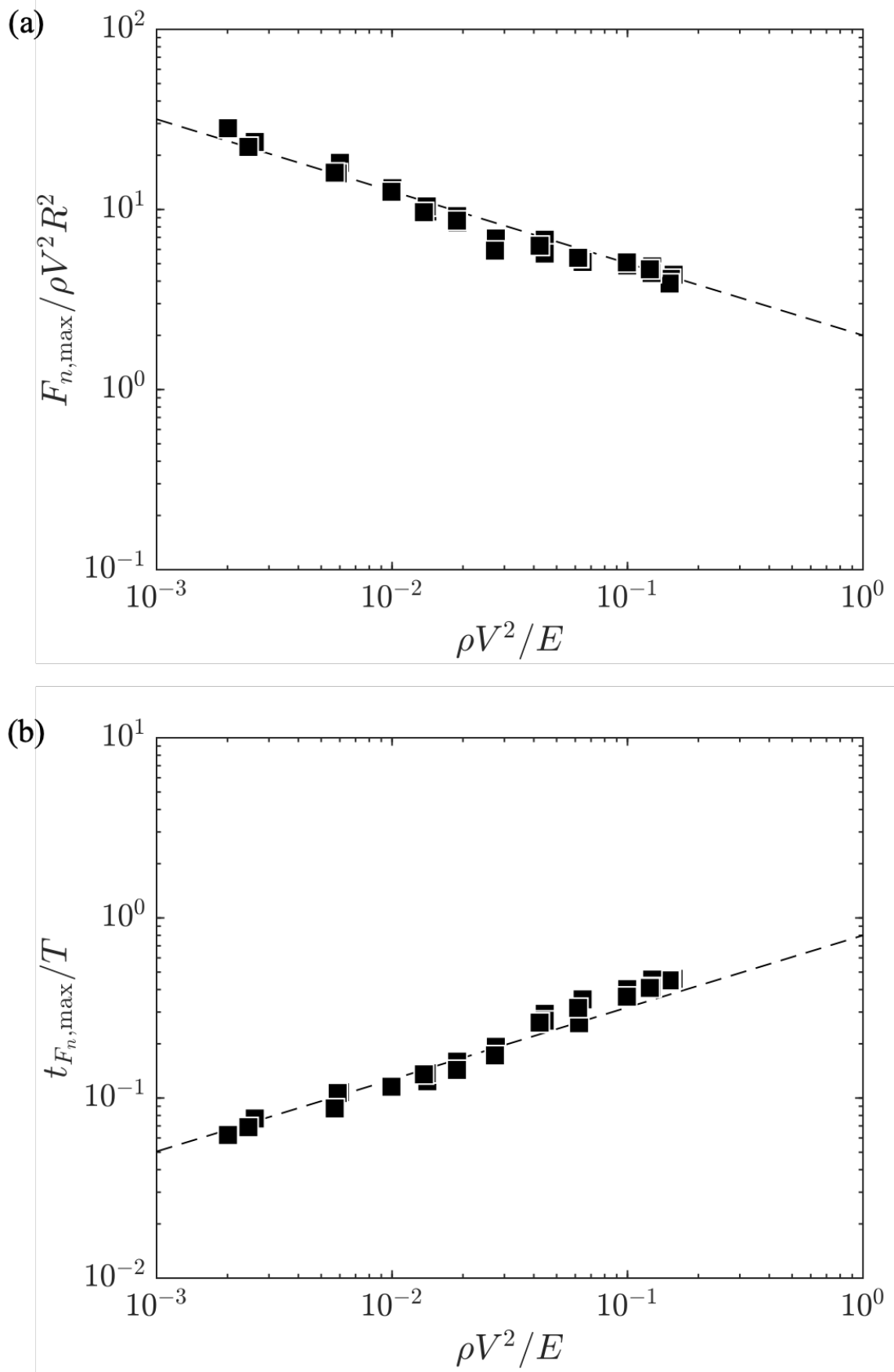}
    \caption{(a) The relationship between the peak force $F_{n, \rm max}$ nondimensionalized by the inertial force $\rho V^2 R^2$ and $\rho V^2/E$. The dashed line shows the scaling law for the peak force (Eq.(\ref{eq:scaling_PeakForce})). (b) The relationship between the peak time $t_{F_n,\rm max}$ nondimensionalized by the impact time $T = R/V$ and $\rho V^2/E$. The dashed line shows the scaling law for the peak times (Eq.(\ref{eq:scaling_PeakTime})).}
    \label{fig:Scaling_ImpactForce}
\end{figure}

Here, we focus on the peak force $F_{n,\rm max}$ and the time when the peak force occurs $t_{F_n,\rm max}$.
The scaling law of the peak force and time with the impact velocity $V$ can be theoretically predicted from the dynamic Hertzian contact model \cite{pradipto2021,kuwabara1987}.
The motion of the equation of a sphere impacting the elastic half-space can be described as
\begin{equation}
    m \frac{d^2z}{dt^2} + F_n = 0, \label{eq:DHCM}
\end{equation}
with initial conditions $dz/dt = V$ and $z = 0$ at $t = 0$, where $z$ is the vertical displacement of the sphere.
Here, we neglect the gravity term because it is negligibly small compared with the other terms.
The scaling law of $F_n$ can be theoretically obtained from the Hertzian contact problem using the displacement of the surface \cite{johnson1985}, which corresponds to $z$ as
\begin{equation}
    F_n = \frac{4}{3}E^\ast R^{1/2}z^{3/2}, \label{eq:Force_HCM}
\end{equation}
where $E^\ast = E/(1-\nu^2)$ is the effective Young's modulus of the gel.
Here, by defining $v = dz/dt$, Eq. (\ref{eq:DHCM}) can be rewritten using Eq. (\ref{eq:Force_HCM}) as
\begin{eqnarray}
    && \frac{4}{3} \pi \rho R^3 v\frac{dv}{dz} + \frac{4}{3}E^\ast R^{1/2} z^{3/2} = 0, \label{eq:DHCM_v}\\
    && \because \, \frac{d^2z}{dt^2} = \frac{dv}{dz}\frac{dz}{dt}. \nonumber
\end{eqnarray}
Thus, Eq. (\ref{eq:DHCM_v}) can be solved for $v$ in the following equation.
\begin{equation}
    v^2 = V^2 - \frac{4}{5}\frac{E^\ast}{\pi \rho} R^{-5/2} z^{5/2} \label{eq:v}
\end{equation}
Because $v = 0$ when $z$ reaches its maximum $z_{\rm max}$,
\begin{equation}\label{eq:scaling z_max}
    z_{\rm max} \propto {E}^{-2/5} \rho^{2/5}  R V^{4/5}.
\end{equation}
By substituting this equation into Eq. (\ref{eq:Force_HCM}), the scaling law for the peak force $F_{n, \rm max}$ can be written as
\begin{equation}
    F_{n, \rm max} \propto {E}^{2/5} \rho^{3/5} R^2 V^{6/5}.
\end{equation}
The peak force $F_{n, \rm max}$ can be nondimensionalized using the inertial force, $\rho V^2 R^2$, as
\begin{equation}
    \frac{F_{n, \rm max}}{\rho V^2 R^2} \propto \left(\frac{\rho V^2}{E}\right)^{-2/5}. \label{eq:scaling_PeakForce}
\end{equation}
Figure \ref{fig:Scaling_ImpactForce}(a) shows the relationship between $F_{n, \rm max}/\rho V^2 R^2$ and $\rho V^2/E$ with Eq. (\ref{eq:scaling_PeakForce}) shown as a dashed line.
The trend of the experimental peak force agrees with the theoretical prediction.

Additionally, we can obtain the scaling law for the peak time $t_{F_n,\rm max}$ from the relationship $z_{\rm max}\propto t_{F_n,\rm max} V$ and Eq. (\ref{eq:scaling z_max}), as
\begin{equation}
    t_{F_n, \rm max} \propto {E}^{-2/5} \rho^{2/5} R V^{-1/5}.
\end{equation}
Let this equation be nondimensionalized by the impact time, $T = R/V$, as
\begin{equation}
    \frac{t_{F_n,\rm max}}{T} \propto \left(\frac{\rho V^2}{E}\right)^{2/5}. \label{eq:scaling_PeakTime}
\end{equation}
Figure \ref{fig:Scaling_ImpactForce}(b) shows the relationship between $t_{F_n, \rm max}/T$ and $\rho V^2/E$ with Eq. (\ref{eq:scaling_PeakTime}) as a dashed line.
The trend of the experimental peak time agrees with the theoretical prediction.

Note that Hertzian contact theory, including the dynamic cases, basically assumes a small deformation of the substrate surface.
Nevertheless, the theoretical predictions of the scaling law still agree well with the experimental data over a wide range of $\rho V^2 /E$ from $\mathcal{O}(10^{-3})$ to $\mathcal{O}(10^{-1})$, even with large substrate deformations.

\section{Conclusion}\label{sec:conclusion}

In this study, all stress components in the dynamic axisymmetric fields of a soft material have been reconstructed using photoelastic tomography.
We have quantitatively validated our approach, focusing on static and dynamic Hertzian contact.

For static Hertzian contact, in which a rigid sphere is pressed against a gel block, all stress tensor components were reconstructed from the photoelastic parameters measured using a polarization camera.
The reconstructed stress fields were compared with the theoretical solution for the stress fields obtained by solving the Hertzian contact problem.
The experimental and theoretical results showed reasonable agreement, indicating that photoelastic tomography can be applied to measuring stress fields in soft materials.
Reconstruction of all stress components allows, for example, analysis of the von Mises stress, which is used for the evaluation of the yield value of a material.

To validate the approach for dynamic Hertzian contact, in which a rigid sphere impacts a gel, the dynamic stress field in the gel was reconstructed from the photoelastic parameters measured.
Consequently, a wide range of stress fields from $\mathcal{O}(10^{-1})$ to $\mathcal{O}(10^1)$ kPa were measured simultaneously with large deformation.
The current measurement system can also stress wave propagation, in which the stress tensor components change rapidly with time.
In addition, the normal impact force acting on the surface was calculated from the reconstructed stress field, and the scaling law for its maximum value was compared with the theoretical prediction for dynamic Hertzian contact.
The experimental results showed good agreement with the theoretically predicted scaling laws, confirming the accuracy of high-speed photoelastic tomography.

We believe that this study has provided an optical methodology for measuring dynamic axisymmetric stress fields in soft materials.
This could lead to a better understanding of phenomena related to various engineering processes, such as the stress distributions in materials caused not only by the impact of rigid spheres but also by the impact of droplets or liquid jets.

\section*{Acknowledgment}

This work was supported by JSPS KAKENHI Grant Numbers JP20H00223, JP22J13343, JP22KJ1239, JP23KJ0859, and JST PRESTO Grant Number JPMJPR21O5, Japan.

\bibliographystyle{elsarticle-num}
\bibliography{ref}

\begin{thebibliography}{10}
\expandafter\ifx\csname url\endcsname\relax
  \def\url#1{\texttt{#1}}\fi
\expandafter\ifx\csname urlprefix\endcsname\relax\def\urlprefix{URL }\fi
\expandafter\ifx\csname href\endcsname\relax
  \def\href#1#2{#2} \def\path#1{#1}\fi

\bibitem{shojima2004}
M.~Shojima, M.~Oshima, K.~Takagi, R.~Torii, M.~Hayakawa, K.~Katada, A.~Morita, T.~Kirino, Magnitude and role of wall shear stress on cerebral aneurysm, Stroke 35~(11) (2004) 2500--2505.
\newblock \href {https://doi.org/10.1161/01.STR.0000144648.89172.0f} {\path{doi:10.1161/01.STR.0000144648.89172.0f}}.

\bibitem{meng2014}
H.~Meng, V.~M. Tutino, J.~Xiang, A.~Siddiqui, High {{WSS}} or low {{WSS}}? {{Complex}} interactions of hemodynamics with intracranial aneurysm initiation, growth, and rupture: toward a unifying hypothesis, American Journal of Neuroradiology 35~(7) (2014) 1254--1262.
\newblock \href {https://doi.org/10.3174/ajnr.A3558} {\path{doi:10.3174/ajnr.A3558}}.

\bibitem{vanooij2015}
P.~{van Ooij}, W.~V. Potters, J.~Collins, M.~Carr, J.~Carr, S.~C. Malaisrie, P.~W.~M. Fedak, P.~M. McCarthy, M.~Markl, A.~J. Barker, Characterization of abnormal wall shear stress using {{4D}} flow {{MRI}} in human bicuspid aortopathy, Annals of Biomedical Engineering 43~(6) (2015) 1385--1397.
\newblock \href {https://doi.org/10.1007/s10439-014-1092-7} {\path{doi:10.1007/s10439-014-1092-7}}.

\bibitem{tagawa2013a}
Y.~Tagawa, N.~Oudalov, A.~E. Ghalbzouri, C.~Sun, D.~Lohse, Needle-free injection into skin and soft matter with highly focused microjets, Lab on a Chip 13~(7) (2013) 1357--1363.
\newblock \href {https://doi.org/10.1039/C2LC41204G} {\path{doi:10.1039/C2LC41204G}}.

\bibitem{kiyama2019a}
A.~Kiyama, N.~Endo, S.~Kawamoto, C.~Katsuta, K.~Oida, A.~Tanaka, Y.~Tagawa, Visualization of penetration of a high-speed focused microjet into gel and animal skin, Journal of Visualization 22~(3) (2019) 449--457.
\newblock \href {https://doi.org/10.1007/s12650-019-00547-8} {\path{doi:10.1007/s12650-019-00547-8}}.

\bibitem{mitchell2019}
B.~R. Mitchell, J.~C. Klewicki, Y.~P. Korkolis, B.~L. Kinsey, The transient force profile of low-speed droplet impact: measurements and model, Journal of Fluid Mechanics 867 (2019) 300--322.
\newblock \href {https://doi.org/10.1017/jfm.2019.141} {\path{doi:10.1017/jfm.2019.141}}.

\bibitem{sun2022}
T.-P. Sun, F.~{\'A}lvarez-Novoa, K.~Andrade, P.~Guti{\'e}rrez, L.~Gordillo, X.~Cheng, Stress distribution and surface shock wave of drop impact, Nature Communications 13~(1) (2022) 1703.
\newblock \href {https://doi.org/10.1038/s41467-022-29345-x} {\path{doi:10.1038/s41467-022-29345-x}}.

\bibitem{mitchell2019a}
B.~R. Mitchell, J.~C. Klewicki, Y.~P. Korkolis, B.~L. Kinsey, Normal impact force of rayleigh jets, Physical Review Fluids 4~(11) (2019) 113603.
\newblock \href {https://doi.org/10.1103/PhysRevFluids.4.113603} {\path{doi:10.1103/PhysRevFluids.4.113603}}.

\bibitem{gordillo2018}
L.~Gordillo, T.-P. Sun, X.~Cheng, Dynamics of drop impact on solid surfaces: evolution of impact force and self-similar spreading, Journal of Fluid Mechanics 840 (2018) 190--214.
\newblock \href {https://doi.org/10.1017/jfm.2017.901} {\path{doi:10.1017/jfm.2017.901}}.

\bibitem{cheng2022}
X.~Cheng, T.-P. Sun, L.~Gordillo, Drop impact dynamics: impact force and stress distributions, Annual Review of Fluid Mechanics 54~(1) (2022) null.
\newblock \href {https://doi.org/10.1146/annurev-fluid-030321-103941} {\path{doi:10.1146/annurev-fluid-030321-103941}}.

\bibitem{frocht1941}
M.~M. Frocht, Photoelasticity: v.1, {John Wiley \& Sons Inc}, 1941.

\bibitem{aben1993a}
H.~Aben, C.~Guillemet, Photoelasticity of glass, {Springer Berlin Heidelberg}, {Berlin, Heidelberg}, 1993.
\newblock \href {https://doi.org/10.1007/978-3-642-50071-8} {\path{doi:10.1007/978-3-642-50071-8}}.

\bibitem{asai2019}
K.~Asai, S.~Yoshida, A.~Yamada, J.~Matsuoka, A.~Errapart, C.~R. Kurkjian, Micro-photoelastic evaluation of indentation-induced stress in glass, Materials Transactions 60~(8) (2019) 1423--1427.
\newblock \href {https://doi.org/10.2320/matertrans.MD201903} {\path{doi:10.2320/matertrans.MD201903}}.

\bibitem{ramesh2016}
K.~Ramesh, V.~Ramakrishnan, Digital photoelasticity of glass: a comprehensive review, Optics and Lasers in Engineering 87 (2016) 59--74.
\newblock \href {https://doi.org/10.1016/j.optlaseng.2016.03.017} {\path{doi:10.1016/j.optlaseng.2016.03.017}}.

\bibitem{ramesh2021}
K.~Ramesh, Developments in photoelasticity, {IOP Publishing}, 2021.
\newblock \href {https://doi.org/10.1088/978-0-7503-2472-4} {\path{doi:10.1088/978-0-7503-2472-4}}.

\bibitem{yokoyama2023}
Y.~Yokoyama, B.~R. Mitchell, A.~Nassiri, B.~L. Kinsey, Y.~P. Korkolis, Y.~Tagawa, Integrated photoelasticity in a soft material: phase retardation, azimuthal angle, and stress-optic coefficient, Optics and Lasers in Engineering 161 (2023) 107335.
\newblock \href {https://doi.org/10.1016/j.optlaseng.2022.107335} {\path{doi:10.1016/j.optlaseng.2022.107335}}.

\bibitem{hall2012}
M.~S. Hall, R.~Long, C.-Y. Hui, M.~Wu, Mapping three-dimensional stress and strain fields within a soft hydrogel using a fluorescence microscope, Biophysical Journal 102~(10) (2012) 2241--2250.
\newblock \href {https://doi.org/10.1016/j.bpj.2012.04.014} {\path{doi:10.1016/j.bpj.2012.04.014}}.

\bibitem{aben1993}
H.~Aben, C.~Guillemet, Integrated photoelasticity, in: H.~Aben, C.~Guillemet (Eds.), Photoelasticity of {{Glass}}, {Springer}, {Berlin, Heidelberg}, 1993, pp. 86--101.
\newblock \href {https://doi.org/10.1007/978-3-642-50071-8_6} {\path{doi:10.1007/978-3-642-50071-8_6}}.

\bibitem{aben1989}
H.~K. Aben, J.~I. Josepson, K.~J.~E. Kell, The case of weak birefringence in integrated photoelasticity, Optics and Lasers in Engineering 11~(3) (1989) 145--157.
\newblock \href {https://doi.org/10.1016/0143-8166(89)90029-8} {\path{doi:10.1016/0143-8166(89)90029-8}}.

\bibitem{aben2000}
H.~Aben, L.~Ainola, J.~Anton, Integrated photoelasticity for nondestructive residual stress measurement in glass, Optics and Lasers in Engineering 33~(1) (2000) 49--64.
\newblock \href {https://doi.org/10.1016/S0143-8166(00)00018-X} {\path{doi:10.1016/S0143-8166(00)00018-X}}.

\bibitem{nelson2021}
D.~V. Nelson, Residual stress determination using full-field optical methods, Journal of Physics: Photonics 3~(4) (2021) 044003.
\newblock \href {https://doi.org/10.1088/2515-7647/ac1ceb} {\path{doi:10.1088/2515-7647/ac1ceb}}.

\bibitem{yoshida2012}
S.~Yoshida, S.~Iwata, T.~Sugawara, Y.~Miura, J.~Matsuoka, A.~Errapart, C.~R. Kurkjian, Elastic and residual stresses around ball indentations on glasses using a micro-photoelastic technique, Journal of Non-Crystalline Solids 358~(24) (2012) 3465--3472.
\newblock \href {https://doi.org/10.1016/j.jnoncrysol.2012.01.069} {\path{doi:10.1016/j.jnoncrysol.2012.01.069}}.

\bibitem{mitchell2023}
B.~Mitchell, Y.~Yokoyama, A.~Nassiri, Y.~Tagawa, Y.~P. Korkolis, B.~L. Kinsey, An investigation of hertzian contact in soft materials using photoelastic tomography, Journal of the Mechanics and Physics of Solids 171 (2023) 105164.
\newblock \href {https://doi.org/10.1016/j.jmps.2022.105164} {\path{doi:10.1016/j.jmps.2022.105164}}.

\bibitem{aben2012}
H.~Aben, A.~Errapart, Photoelastic tomography with linear and non-linear algorithms, Experimental Mechanics 52~(8) (2012) 1179--1193.
\newblock \href {https://doi.org/10.1007/s11340-011-9575-z} {\path{doi:10.1007/s11340-011-9575-z}}.

\bibitem{aben2010a}
H.~Aben, L.~Ainola, A.~Errapart, Photoelastic tomography as hybrid mechanics, EPJ Web of Conferences 6 (2010) 32009.
\newblock \href {https://doi.org/10.1051/epjconf/20100632009} {\path{doi:10.1051/epjconf/20100632009}}.

\bibitem{aben2008}
H.~Aben, J.~Anton, A.~Errapart, Modern photoelasticity for residual stress measurement in glass, Strain 44~(1) (2008) 40--48.
\newblock \href {https://doi.org/10.1111/j.1475-1305.2008.00422.x} {\path{doi:10.1111/j.1475-1305.2008.00422.x}}.

\bibitem{errapart2011}
A.~Errapart, Determination of all stress components of axisymmetric stress state in photoelastic tomography, Applied Mechanics and Materials 70 (2011) 434--439.
\newblock \href {https://doi.org/10.4028/www.scientific.net/AMM.70.434} {\path{doi:10.4028/www.scientific.net/AMM.70.434}}.

\bibitem{anton2008}
J.~Anton, A.~Errapart, H.~Aben, L.~Ainola, A discrete algorithm of integrated photoelasticity for axisymmetric problems, Experimental Mechanics 48~(5) (2008) 613--620.
\newblock \href {https://doi.org/10.1007/s11340-008-9121-9} {\path{doi:10.1007/s11340-008-9121-9}}.

\bibitem{yoneyama2006}
S.~Yoneyama, H.~Kikuta, K.~Moriwaki, Simultaneous observation of phase-stepped photoelastic fringes using a pixelated microretarder array, Optical Engineering 45~(8) (2006) 083604.
\newblock \href {https://doi.org/10.1117/1.2335894} {\path{doi:10.1117/1.2335894}}.

\bibitem{onuma2014}
T.~Onuma, Y.~Otani, A development of two-dimensional birefringence distribution measurement system with a sampling rate of 1.3mhz, Optics Communications 315 (2014) 69--73.
\newblock \href {https://doi.org/10.1016/j.optcom.2013.10.086} {\path{doi:10.1016/j.optcom.2013.10.086}}.

\bibitem{miyazaki2021}
Y.~Miyazaki, M.~Usawa, S.~Kawai, J.~Yee, M.~Muto, Y.~Tagawa, Dynamic mechanical interaction between injection liquid and human tissue simulant induced by needle-free injection of a highly focused microjet, Scientific Reports 11~(1) (2021) 14544.
\newblock \href {https://doi.org/10.1038/s41598-021-94018-6} {\path{doi:10.1038/s41598-021-94018-6}}.

\bibitem{johnson1985}
K.~L. Johnson, Contact mechanics, {Cambridge University Press}, {Cambridge}, 1985.
\newblock \href {https://doi.org/10.1017/CBO9781139171731} {\path{doi:10.1017/CBO9781139171731}}.

\bibitem{ike2019}
C.~C. Ike, Love stress function method for solving axisymmetric elasticity problems of the elastic half-space, Electronic Journal of Geotechnical Engineering 24~(3) (2019) 44.

\bibitem{love1929}
A.~E.~H. Love, {{IX}}. the stress produced in a semi-infinite solid by pressure on part of the boundary, Philosophical Transactions of the Royal Society of London. Series A, Containing Papers of a Mathematical or Physical Character 228~(659-669) (1929) 377--420.
\newblock \href {https://doi.org/10.1098/rsta.1929.0009} {\path{doi:10.1098/rsta.1929.0009}}.

\bibitem{pradipto2021}
{Pradipto}, H.~Hayakawa, Impact-induced hardening in dense frictional suspensions, Physical Review Fluids 6~(3) (2021) 033301.
\newblock \href {https://doi.org/10.1103/PhysRevFluids.6.033301} {\path{doi:10.1103/PhysRevFluids.6.033301}}.

\bibitem{kuwabara1987}
G.~Kuwabara, K.~Kono, Restitution coefficient in a collision between two spheres, Japanese Journal of Applied Physics 26~(8R) (1987) 1230.
\newblock \href {https://doi.org/10.1143/JJAP.26.1230} {\path{doi:10.1143/JJAP.26.1230}}.

\bibitem{howland2016}
C.~J. Howland, A.~Antkowiak, J.~R. {Castrej{\'o}n-Pita}, S.~D. Howison, J.~M. Oliver, R.~W. Style, A.~A. {Castrej{\'o}n-Pita}, It's harder to splash on soft solids, Physical Review Letters 117~(18) (2016) 184502.
\newblock \href {https://doi.org/10.1103/PhysRevLett.117.184502} {\path{doi:10.1103/PhysRevLett.117.184502}}.

\bibitem{basso2020a}
B.~C. Basso, J.~B. Bostwick, Splashing on soft elastic substrates, Langmuir 36~(49) (2020) 15010--15017.
\newblock \href {https://doi.org/10.1021/acs.langmuir.0c02500} {\path{doi:10.1021/acs.langmuir.0c02500}}.

\bibitem{rapet2019}
J.~Rapet, Y.~Tagawa, C.~D. Ohl, Shear-wave generation from cavitation in soft solids, Applied Physics Letters 114~(12) (2019) 123702.
\newblock \href {https://doi.org/10.1063/1.5083141} {\path{doi:10.1063/1.5083141}}.

\bibitem{aben1997}
H.~Aben, A.~Puro, Photoelastic tomography for three-dimensional flow birefringence studies, Inverse Problems 13~(2) (1997) 215--221.
\newblock \href {https://doi.org/10.1088/0266-5611/13/2/002} {\path{doi:10.1088/0266-5611/13/2/002}}.

\bibitem{doyle1982}
J.~F. Doyle, On a nonlinearity in flow birefringence, Experimental Mechanics 22~(1) (1982) 37--38.
\newblock \href {https://doi.org/10.1007/BF02325702} {\path{doi:10.1007/BF02325702}}.

\bibitem{aben2010}
H.~Aben, L.~Ainola, A.~Errapart, Application of the abel inversion in case of a tensor field, Inverse Problems in Science and Engineering 18~(2) (2010) 241--249.
\newblock \href {https://doi.org/10.1080/17415970903545124} {\path{doi:10.1080/17415970903545124}}.

\bibitem{aben1992}
H.~K. Aben, S.~J. Idnurm, J.~Josepson, K.-J.~E. Kell, A.~E. Puro, Optical tomography of the stress tensor field, in: Analytical {{Methods}} for {{Optical Tomography}}, Vol. 1843, {SPIE}, 1992, pp. 220--229.
\newblock \href {https://doi.org/10.1117/12.131894} {\path{doi:10.1117/12.131894}}.

\bibitem{dasch1992}
C.~J. Dasch, One-dimensional tomography: a comparison of abel, onion-peeling, and filtered backprojection methods, Applied Optics 31~(8) (1992) 1146--1152.
\newblock \href {https://doi.org/10.1364/AO.31.001146} {\path{doi:10.1364/AO.31.001146}}.

\bibitem{xiong2020a}
Y.~Xiong, T.~Kaufmann, N.~Noiray, Towards robust {{BOS}} measurements for axisymmetric flows, Experiments in Fluids 61~(8) (2020) 178.
\newblock \href {https://doi.org/10.1007/s00348-020-03007-4} {\path{doi:10.1007/s00348-020-03007-4}}.

\bibitem{ichihara2022}
S.~Ichihara, T.~Shimazaki, Y.~Tagawa, Background-oriented schlieren technique with vector tomography for measurement of axisymmetric pressure fields of laser-induced underwater shock waves, Experiments in Fluids 63~(11) (2022) 182.
\newblock \href {https://doi.org/10.1007/s00348-022-03524-4} {\path{doi:10.1007/s00348-022-03524-4}}.

\bibitem{popov2017}
V.~L. Popov, Contact mechanics and friction: physical principles and applications, 2nd Edition, {Springer-Verlag}, {Berlin Heidelberg}, 2017.
\newblock \href {https://doi.org/10.1007/978-3-662-53081-8} {\path{doi:10.1007/978-3-662-53081-8}}.

\bibitem{mitchell2022}
B.~Mitchell, Water droplet machining and droplet impact mechanics, Ph.D. thesis, University of New Hampsiher (May 2022).

\bibitem{su2023a}
F.~Su, Z.~Wang, Error analysis and correction of a photoelastic method based on a pixelated polarization camera, Optics and Lasers in Engineering 161 (2023) 107374.
\newblock \href {https://doi.org/10.1016/j.optlaseng.2022.107374} {\path{doi:10.1016/j.optlaseng.2022.107374}}.

\bibitem{lane2022a}
C.~Lane, D.~Rode, T.~R{\"o}sgen, Calibration of a polarization image sensor and investigation of influencing factors, Applied Optics 61~(6) (2022) C37--C45.
\newblock \href {https://doi.org/10.1364/AO.437391} {\path{doi:10.1364/AO.437391}}.

\bibitem{otani1994}
Y.~Otani, T.~Shimada, T.~Yoshizawa, N.~Umeda, Two-dimensional birefringence measurement using the phase shifting technique, Optical Engineering 33~(5) (1994) 1604--1609.
\newblock \href {https://doi.org/10.1117/12.168435} {\path{doi:10.1117/12.168435}}.

\bibitem{onuma2012}
T.~Onuma, Y.~Otani, A dynamic measurement system for two-dimensional birefringence distribution with sub-millisecond time resolution, Journal of the Japan Society for Precision Engineering 78~(12) (2012) 1082--1086.
\newblock \href {https://doi.org/10.2493/jjspe.78.1082} {\path{doi:10.2493/jjspe.78.1082}}.

\bibitem{harris1978}
J.~K. Harris, A photoelastic substrate technique for dynamic measurements of forces exerted by moving organisms, Journal of Microscopy 114~(2) (1978) 219--228.
\newblock \href {https://doi.org/10.1111/j.1365-2818.1978.tb00132.x} {\path{doi:10.1111/j.1365-2818.1978.tb00132.x}}.

\bibitem{bayley1959}
H.~G. Bayley, Gelatin as a photo-elastic material, Nature 183~(4677) (1959) 1757--1758.
\newblock \href {https://doi.org/10.1038/1831757a0} {\path{doi:10.1038/1831757a0}}.

\bibitem{zhang2024}
H.~Zhang, S.~Jia, B.~Wen, Z.~Yang, X.~Zhou, Z.~Lin, L.~Wang, Advancing instantaneous photoelastic method with color polarization camera, Optics and Lasers in Engineering 172 (2024) 107868.
\newblock \href {https://doi.org/10.1016/j.optlaseng.2023.107868} {\path{doi:10.1016/j.optlaseng.2023.107868}}.

\end{thebibliography}

\end{document}